\documentclass{article}

\PassOptionsToPackage{comma,compress,round}{natbib}
\usepackage[preprint]{neurips_2025}

\usepackage[utf8]{inputenc} 
\usepackage[T1]{fontenc}    
\usepackage{nicefrac}       
\usepackage{microtype}      
\usepackage{xcolor}         
\usepackage{graphicx,verbatim}
\usepackage{amsmath}
\usepackage{amssymb}
\usepackage{arydshln}
\usepackage{bm}
\usepackage{booktabs,makecell,multirow,pifont,color, colortbl}
\usepackage{array}
\usepackage{tabularx} 
\newcommand{\cmark}{\ding{51}} 

\usepackage[table]{xcolor} 
\usepackage{makecell}
\definecolor{lightmaroon}{RGB}{200,120,120}
\usepackage{makecell,xcolor}

\usepackage{hyperref}
\usepackage{url}
\usepackage{float}
\usepackage{wrapfig}
\usepackage{enumitem} 
\definecolor{lightmaroon}{RGB}{200,120,120}
\usepackage{soul,xcolor}
\sethlcolor{lightmaroon!20}
\usepackage{makecell,xcolor}

\definecolor{CBBlue}{HTML}{56B4E9}   
\definecolor{CBGreen}{HTML}{009E73}  

\definecolor{first}{rgb}{1, 0.7, 0.7}
\definecolor{second}{rgb}{1, 0.85, 0.7}
\definecolor{lightblue}{rgb}{0.68,0.85,0.90}
\newcommand{\best}[1]{\cellcolor{CBBlue!28}\textbf{#1}}   
\newcommand{\ssbest}[1]{\cellcolor{CBGreen!22}#1}         

\newcommand{\sbest}[1]{\underline{#1}}

\definecolor{hlb}{HTML}{ADD8E6} 
\DeclareRobustCommand{\hlblue}[1]{{\sethlcolor{blue!10}\hl{#1}}}

\DeclareRobustCommand{\hlb}[1]{{\sethlcolor{lightmaroon!25}\hl{#1}}}
\DeclareRobustCommand{\hlgreen}[1]{{\sethlcolor{CBGreen!22}\hl{#1}}}

\title{Mitigating Surgical Data Imbalance with Dual-Prediction Video Diffusion Model}

\author{%
  Danush Kumar Venkatesh$^{1,a}$ \quad Adam Schmidt$^{2,b}$ \quad Muhammad Abdullah Jamal$^{b}$ \quad Omid Mohareri$^{b}$\\
  $^{a}$Department of Translational Surgical Oncology, NCT/UCC Dresden, a partnership between DKFZ, \\
  Faculty of Medicine and University Hospital Carl Gustav Carus, TUD Dresden,HZDR, Germany \\
  $^{b}$Intuitive Surgical, Inc., Sunnyvale, CA, United States\\
}

\begin{document}

\maketitle

\begin{abstract}
  Surgical video datasets are essential for scene understanding, enabling procedural modeling and intra-operative support. However, these datasets are often heavily imbalanced, with rare actions and tools under-represented, which limits the robustness of downstream models. We address this challenge with \emph{SurgiFlowVid}, a sparse and controllable video diffusion framework for generating surgical videos of under-represented classes. Our approach introduces a dual-prediction diffusion module that jointly denoises RGB frames and optical flow, providing temporal inductive biases to improve motion modeling from limited samples. In addition, a sparse visual encoder conditions the generation process on lightweight signals (e.g., sparse segmentation masks or RGB frames), enabling controllability without dense annotations. We validate our approach on three surgical datasets across tasks including action recognition, tool presence detection, and laparoscope motion prediction. Synthetic data generated by our method yields consistent gains of $10$–$20\%$ over competitive baselines, establishing \emph{SurgiFlowVid} as a promising strategy to mitigate data imbalance and advance surgical video understanding methods.
\end{abstract}

\section{Introduction}\label{into}


Robotic-assisted minimally invasive surgery (RAMIS) has become a cornerstone of modern surgical practice, offering patients reduced trauma, faster recovery, and fewer complications~\citep{steffi_robot, robot_2016}. However, operating using an endoscopic video feed rather than direct vision introduces challenges such as limited depth perception, reduced haptic feedback, and altered hand–eye coordination. These limitations increase both the cognitive and technical demands placed on surgeons during procedures~\citep{sorensen2016three, robot_2024}.
\footnote{Work done during an internship at Intuitive Surgical Inc.}
\footnote{Corresponding author: Adam Schmidt, Adam.Schmidt@intusurg.com}

\begin{figure}[h]
\begin{center}
\includegraphics[width=\textwidth,keepaspectratio]{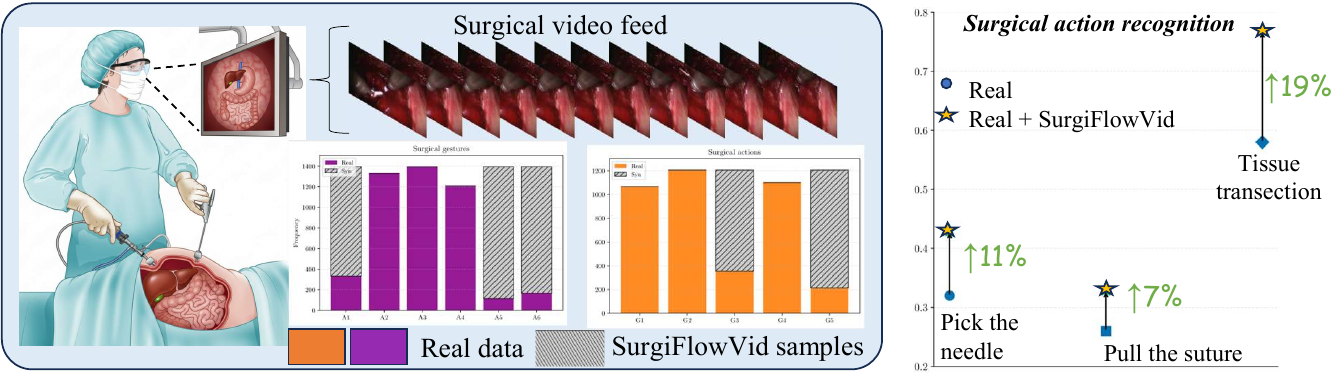}
\end{center}
\caption{\textbf{Data challenge in the surgical domain}. During a laparoscopic procedure, the surgeon operates via the endoscopic video feed (video on the monitor). ML models can leverage these videos for providing guidance through surgical scene understanding. However, the datasets are \emph{skewed} as shown in the bar plots. We aim to mitigate data imbalance with synthetic samples. The right plot shows improvements from adding samples generated from our approach (SurgiFlowVid).} 
\label{fig:intro}
\end{figure}

The emerging field of \emph{Surgical Data Science} seeks to address these challenges by developing computational methods that leverage the video data generated during surgery. In particular, deep learning (DL) methods could be utilized to understand the surgical scene, thereby supporting intraoperative decisions and reducing the burden on surgeons. Surgical video datasets, therefore, play a central role in enabling tasks, including surgical phase and gesture recognition~\citep{padoy2012statistical, funke2025tunes, funke2019using}, instrument detection and segmentation~\citep{nwoye2022rendezvous,kolbinger2023anatomy}, tool tracking~\citep{schmidt2024tracking}, and skill assessment~\citep{funke2019video, hoffmann2024aixsuture}. However, despite recent efforts to release annotated datasets~\citep{gynsurg, grasp, sar, autolaparo}, these resources remain heavily imbalanced, with rare actions, steps, or tool usages under-represented (see Fig.~\ref{fig:intro}). Such skewed distributions limit the generalization of DL models. Common approaches such as class-sampling and augmentation can increase the frequency of these samples but do not contribute to the diversity of the dataset.

The data imbalance challenge in surgical datasets have motivated increasing interest in synthetic data generation. With the advent of diffusion models (DMs)~\citep{ho2020denoising,dhariwal2021diffusion}, synthetic surgical images have been successfully utilized to augment real datasets, thereby reducing imbalance and enhancing downstream performance~\cite{venkatesh2025data,frisch2023synthesising,nwoye2025surgical}. However, extending DMs to surgical video generation remains underexplored due to the substantial demands in data and compute. While recent progress in video synthesis is promising, controllability is especially critical in the surgical domain, where specific tools and anatomical structures must appear in procedure-dependent contexts (e.g., laparoscopy vs. robotic surgery). Prior work often relies on dense per-frame segmentation masks to control video generation~\citep{hierasurg,sg2vid,visage,iliash2024interactive,surgen}, but these require costly expert annotations that are rarely available. In practice, surgical datasets typically contain only sparse segmentation masks—or none at all—while under-represented classes are particularly scarce. This raises a critical question: \textit{how can generative models improve learning for under-represented classes when only sparse or no conditional signals are available?}

To address this challenge, we propose \emph{SurgiFlowVid} (\textbf{Surgi}cal \textbf{Flow}-inducted \textbf{Vid}eo generation), a diffusion-based framework designed to synthesize spatially and temporally coherent surgical videos of under-represented classes. 
We introduce a dual-prediction approach that jointly denoises RGB frames and optical flow maps, providing inductive biases to improve motion modeling from limited data. Beyond text prompts, SurgiFlowVid can condition directly on RGB frames or sparse segmentation masks, when available,  via a visual encoder. While video DMs typically rely on heavy compute, our approach is tailored to the constrained settings common in healthcare, ensuring practical applicability.
SurgiFlowVid generates diverse and coherent videos of under-represented classes, which we use to augment real datasets and evaluate the models across multiple datasets and downstream tasks.
By tackling the challenges of data imbalance, our approach advances robust DL methods for surgical video understanding, contributing to the broader goal of improving surgical healthcare. 
We summarize our contributions as follows:
\begin{enumerate}
\item We address the critical challenge of data imbalance in surgical datasets by synthesizing video samples of under-represented classes with diffusion models, providing a principled way to augment real world datasets.  
\item We introduce \emph{SurgiFlowVid}, a surgical video diffusion framework equipped with a dual-prediction diffusion U-Net that leverages both RGB frames and optical flow to capture spatio-temporal relationships, even in the minimal available video samples of under-represented classes. In addition, a visual encoder enables conditioning on sparse conditional frames when available, removing the need for costly dense annotations.
\item We extensively evaluate the proposed framework, starting with an analysis of synthetic data attributes and extending to three surgical datasets across diverse surgical downstream tasks: action recognition, tool presence detection, and laparoscope motion prediction. The results show consistent performance gains of $10$–$20\%$ over strong baselines, highlighting the effectiveness of our approach in advancing robust surgical video understanding models.
\end{enumerate}


\section{Related Work}\label{rel_work}
\paragraph{Synthetic data in surgery} $2$D synthetic laparoscopic surgical images generated using
GANs~\citep{gan} and diffusion models (DMs)~\citep{diffusion1,difusion2} have been shown to enhance downstream tasks~\citep{venkatesh2024exploring,venkatesh2025data,frisch2023synthesising,nwoye2025surgical,allmendinger2024navigating,simuscope, micha}. 
However, these approaches remain limited to static image generation and fail to capture the temporal context essential for surgical videos, which are the primary data source in real-time procedures. While diffusion models have also shown success in medical imaging domains such as MRI and CT~\citep{dorjsembe2022three,khader2023denoising,zhao2025maisi}, these modalities differ fundamentally from surgical video data.

\paragraph{Surgical Video Synthesis} Although laparoscopic video synthesis has attracted increasing attention in recent years, its potential for addressing data imbalance in surgical tasks remains underexplored. Endora~\citep{endora} introduced unconditional video generation by incorporating semantic features from a DINO~\citep{dino} backbone, while
MedSora~\citep{medsora} proposed a framework based on a Mamba diffusion model. However, both approaches lacked controllability, which is crucial for generating task-specific videos that can mitigate data imbalance. ~\citet{iliash2024interactive} and SurGen~\citep{surgen} extended video generation by conditioning on pre-defined instrument masks to synthesize coherent surgical phases. Yet, these methods requires vast quantities of labeled real data ($\approx200$K videos), which restricts its applicability to well-studied procedures, such as cholecystectomy~\citep{nwoye2021rendezvous,twinanda2016endonet}, and prevents its generalization to less documented surgeries.

Other works, such as VISAGE~\citep{visage} and SG2VID~\citealp{sg2vid}, condition generation on action graphs
which require curated datasets with detailed annotations and they are often unavailable for many surgical procedures. SurgSora~\citep{surgsora} instead conditions video synthesis on user-defined instrument trajectories, whereas Bora~\citep{bora} leverages large language models (LLMs) to generate instruction prompts for controlling video generation. 
More recently, SurV-Gen~\citep{venkatesh2025mission} was proposed as a video diffusion framework for generating samples of rare classes. This method 
employs a rejection sampling strategy to filter out degenerate cases (poor consistency) of synthetic videos from a large candidate pool. Although there exists plethora of state-of-the-art video diffusion models for the natural domain~\citep{svd,cogvideox,cosmos,moviegen}, adapting them for the surgical domain is challenging due to the large amounts of curated video data and compute needed to train them. Additional related work is in the appendix (\ref{sec:add_related_work}).
%

Our approach, although closely related to SurV-Gen, introduces notable advantages: by incorporating optical flow as an inductive bias, we generate temporally coherent and plausible videos without the need for rejection sampling. Additionally, by conditioning on sparse segmentation masks or RGB frames, we achieve greater controllability and diversity in generating under-represented classes.


\section{SurgiFlowVid}\label{back}
Our goal is to alleviate data imbalance by generating spatially and temporally coherent surgical videos of under-represented classes, a task that is made difficult by the limited data available to model spatial and temporal dynamics accurately. To address this, we introduce \emph{SurgiFlowVid}, which includes a multi-stage conditioninal training process built upon the  SurV-Gen framework~\citep{venkatesh2025mission} with the following core modifications:

(i) \textit{Dual-prediction diffusion U-Net:} we introduce a U-Net module that jointly predicts RGB frames and optical flow maps during training, enabling the model to capture temporal motion alongside spatial appearance which cannot be reliably inferred from RGB appearance alone. 

\begin{figure}[h]
\begin{center}
\includegraphics[height=5cm, width=\textwidth,keepaspectratio]{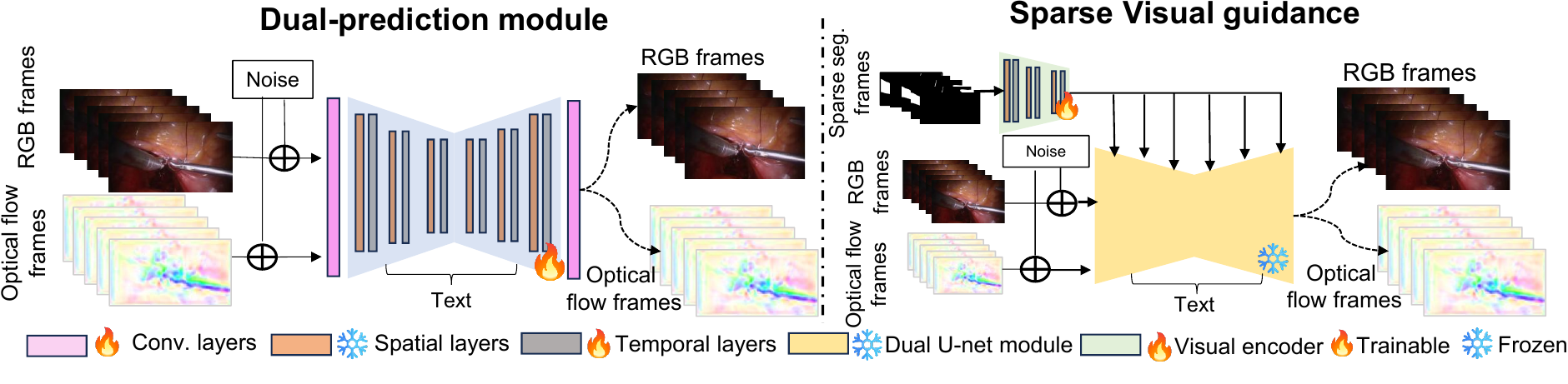}
\end{center}
\caption{\textbf{SurgiFlowVid approach.}The dual-prediction diffusion U-Net module reconstructs both RGB and optical flow frames from noised inputs to capture spatio-temporal dynamics from limited data. Sparse visual encoder is trained with segmentation masks (if available) or RGB frames for conditional generation; optical flow is used only during training.}
\label{fig:model}
\end{figure}


(ii) \textit{Sparse conditional guidance:} dense segmentation masks are rarely available in surgical datasets, and relying solely on text or label conditioning provides weak guidance. Instead, we design a sparse visual guidance encoder that conditions the diffusion process on either the available sparse segmentation masks or the RGB frames from the input video. 
Our model supports both text-based unconditional generation and conditional generation with sparse masks (if available), generating under-represented class samples with spatio-temporal consistency. 

We first review SurV-Gen and follow it with explaining our approach. The overview of our approach is shown in Fig.~\ref{fig:model}.

\paragraph{(i) Surgical Video Generation}\label{sec:survgen}
We build our framework on top of the SurV-Gen model, which follows a two-stage training strategy. In the first stage, Stable Diffusion (SD)~\citep{sd} is adopted as the base text-to-image model, where the diffusion process is performed in the latent space. An image $x_0$ is first encoded into $z_0$ via an encoder $E(x_0)$, and during the forward diffusion process $z_0$ is iteratively perturbed as  $z_t = \sqrt{\bar{\alpha}_t} \, z_0 + \sqrt{1 - \bar{\alpha}_t} \, \epsilon,$ with $\epsilon \sim \mathcal{N}(0,I)$, $\alpha_t = 1 - \beta_t$, and $\bar{\alpha}_t = \prod_{s=0}^t \alpha_s$, where $\beta_t$ determines the noise strength. A denoising network $\epsilon_\theta(\cdot)$ is trained to reverse this process by minimizing the reconstruction loss  

\begin{equation}
    \mathcal{L} = \mathbb{E}_{E(x_0), y, \epsilon \sim \mathcal{N}(0,I), t} 
    \big[ \lVert \epsilon - \epsilon_\theta(z_t, t, y) \rVert_2^2 \big],
\end{equation}

where $y$ denotes the text prompt associated with $x_0$.  In the second stage, the fine-tuned spatial layers of the SD model is extended to operate directly on surgical video sequences. Temporal transformer blocks~\citep{vaswani2017attention} are inserted after each spatial block while keeping spatial layers frozen, thereby focusing the training on temporal dynamics. Given a video tensor $v \in \mathbb{R}^{b \times c \times f \times h \times w}$, where $b$ is the batch size, $f$ the number of frames, $h$, $w$ and $c$ are the height, width and channel dimensions respectively, the temporal layers reshape $v$ to $(bhw) \times f \times c$ and apply self-attention: $v_{\text{out}} = \text{Softmax}\left(\frac{QK^T}{\sqrt{c}}\right)V$
with $Q = v_{\text{in}}W_Q$, $K = v_{\text{in}}W_K$, and $V = v_{\text{in}}W_V$ as the query, key, and value projections. Cross-frame attention captures motion dynamics, but relying solely on it—or on text and label conditioning—is insufficient to model tool and tissue motion. 
\paragraph{(ii) Dual-prediction module}\label{sec:surgi} In our approach, we modify the U-net such that optical flow, $p$, is taken as an input along with input tensor, $v$. Given two consecutive frames $v_1, v_2 \in \mathbb{R}^{3 \times H \times W}$, the optical flow is computed as  $D_t(v_1,v_2)=(d_1,d_2)$, which encodes the pixel displacement at location $(v_1,v_2)$. We convert $D_t$ into an RGB image by computing a normalized magnitude $r(v_1,v_2)$ and angle $\theta$:  
\[
r(v_1,v_2) \;=\;  \frac{\sqrt{\widehat{d}_1^2 + \widehat{d}_2^2}}{\|D_t\|_{\max}+\varepsilon},
\qquad
\theta(v_1,v_2) \;=\; \frac{1}{\pi}\,\mathrm{atan2}\!\big(\widehat{d}_2,\,\widehat{d}_1\big),
\]  
where $\widehat{d}_1, \widehat{d}_2$ denote the normalized flow components and $\varepsilon > 0$ ensures numerical stability. The angle $\theta$ is mapped to a color, while the magnitude $r$ attenuates this color to produce the RGB encoding resulting in the flow tensor $p^{c \times (f-1) \times h \times w}$. We define the \emph{dual-prediction} diffusion U-Net by modifying its input and output layers to process RGB frames and optical flow jointly. Specifically, the first layer is adapted to accommodate both tensors, $v$ and $p$, while the final layer is modified to predict both RGB and flow frames. These layers are trained together with the temporal attention layers using the loss function ($L$) defined as,
\begin{equation}\label{eq:loss}
    \mathcal{L} = \mathbb{E}_{E(x_0), y, \epsilon \sim \mathcal{N}(0,I), t} 
    \Big[ \lVert \epsilon - \epsilon_{\theta}(z_t, t, y) \rVert_2^2 
    + \lambda_p \, \lVert \epsilon - \epsilon_{\theta}(z_p, t, y) \rVert_2^2 \Big],
\end{equation}  
where $z_p$ is the noised optical flow frames and $\lambda_p$ is a weighting parameter. The model jointly denoises each chunk of RGB and flow frames. We freeze the spatial layers in this stage and optical flow is used solely as a training-time signal with text prompts being used during sampling.
\paragraph{(iii) Sparse visual guidance}\label{sec:vis_end} To incorporate conditional guidance, we extend the sparse condition encoder proposed in SparseCtrl~\citep{sparse}, which propagates sparse signals (e.g., frames) across time using spatial and temporal layers to improve consistency between conditioned and unconditioned frames. 
In our framework, we integrate the dual-prediction U-net and redefine the \emph{sparse visual encoder} as a lightweight module that encodes only the sparse conditional frames. The U-Net backbone is frozen, and only the visual encoder is optimized using the loss in Eq.~\ref{eq:loss}. By incorporating optical flow into the loss, we explicitly supervise both motion and structure, allowing the model to move beyond appearance propagation alone, thereby reducing data requirements and improving robustness. Formally, given sparse conditional signals $s_s \in \mathbb{R}^{3 \times H \times W}$ (e.g., RGB frame or segmentation mask) and a binary mask $m \in \{0,1\}^{1 \times H \times W}$ indicating whether a frame is conditioned, the sparse encoder input is constructed as  $\hat{c} = [\,s_s \,\|\, m\,] $  where $\|$ denotes channel-wise concatenation. This design offers flexibility by enabling diverse conditional inputs to guide the generation process. At inference, we sample sparse frames from the real dataset and reassign them to different temporal positions to synthesize videos.  

\section{Experiments}\label{exps}
In this section, we outline our experimental setup, evaluation schemes and the downstream tasks we evaluate the generated synthetic datasets. 
We focus particularly on the under-represented classes, and generate videos of such classes to match their instances to the well represented ones. 

\paragraph{Datasets}
(i) \textbf{SAR-RARP50}
 consists of radical prostatectomy (robotic) videos from $50$ patients, with a split of $35$, $5$, and $10$ patients for training, validation, and test sets, respectively~\citep{sar}. The annotated surgical actions include: picking up the needle (A$1$), positioning the needle (A$2$), pushing the needle through tissue (A$3$), pulling the needle (A$4$), cutting the suture (A$5$), tying the knot (A$6$), and returning the needle (A$7$). Since action A$6$ occurs only once in the test set, it is omitted from evaluation. The under-represented classes in this dataset are A$1$, A$5$, and A$7$.
 The primary task involves recognizing the surgical action at time $t$ given a video clip. In addition, segmentation masks are available for nine classes collected at $1$fps.
 Using these masks, we construct the task of surgical tool presence detection, where the objective is to identify which instruments are present in a given surgical video.  \\
 (ii) \textbf{GraSP}
includes robotic prostatectomy procedures~\citep{grasp}. It consists of $13$ patients with a two-fold cross-validation setup, where five patients are held out for testing. The dataset contains annotations for $20$ different surgical actions. For this study, we focus on a subset of five actions: pulling the suture (G$1$), tying the suture (G$2$), cutting the suture (G$3$), cutting between the prostate and bladder neck (G$4$), and identifying the iliac artery (G$5$). All classes except G$5$ are under-represented. Instrument annotations are also provided for six classes at every $35$ secs making them sparse in nature.
We use this dataset for both surgical action recognition and tool presence detection tasks. \\ 
(iii) \textbf{AutoLaparo} 
contains laparoscopic hysterectomy videos from 21 patients, with annotations describing the movements of the laparoscope~\citep{autolaparo}. In total, it contains approximately $300$ clips, each lasting $10$ seconds, covering six motion types: up, down, left, right, zoom-in, and zoom-out. The laparoscope motion occurs precisely at the $5$th second of each clip, enabling the formulation of two tasks. In the \textit{online} recognition setting, only the first $5$ seconds are provided to the model to predict the upcoming motion, which is particularly relevant for real-time applications. In the \textit{offline} setting, the entire clip is available, and the task is to classify the laparoscope motion using full temporal context. These annotations can be used for developing automatic field-of-view control systems. Owing to the limited dataset size, all movement classes are considered under-represented. 

\paragraph{Baselines} For comparison, we evaluate against recent surgical video diffusion models.  Endora~\citep{endora} is a fully transformer-based unconditional diffusion model, which we train separately on each minor class due to its lack of controllability. SurV-Gen~\citep{venkatesh2025mission} serves as a conditional baseline with both text and label guidance. We also include its rejection sampling (RS) strategy, which filters out degenerate generations and thus represents a strong reference baseline. In addition, we adapt the SparseCtrl~\citep{sparse} model, an effective conditional video diffusion approach that generates videos conditioned on text and sparse conditional masks. We train the SurgiFlowVid model with only text conditioning and follow it by sparse segmentation and RGB frames. These serve as both baselines and ablations of our approach. We maintain a patient specific test split and train the model only on the train split. 
Videos of $16$-frames are generated at four frames-per-second aligning with the requirements of the downstream task.
Together, these baselines span unconditional, conditional, and sparse conditional video diffusion approaches, providing a comprehensive reference for evaluating our method. Additional details are in the appendix (~\ref{sec:model_analysis}). 

\paragraph{Evaluation scheme} We systematically structure our experimental design into three parts to evaluate the role of synthetic data in addressing class imbalance.  

(i) \textbf{Synthetic data attributes:} We analyze which attributes of synthetic data are essential for improving downstream performance. To this end, we conduct controlled experiments on the surgical action recognition task.
First, we \emph{duplicate} the training set and train for the same number of epochs to evaluate whether performance gains arise from true data diversity rather than simple repetition.
Second, to assess the effects of \emph{spatial} and \emph{temporal} consistency, we simulate degraded data by applying elastic deformations and noise to video frames (disrupting spatial structure) and by shuffling frames (disrupting temporal order). Third, we evaluate the effect of \emph{sparse conditioning} by constructing videos from only sparse frames and examining their impact on downstream performance.


(ii) \textbf{Class modeling:} We investigate whether synthetic data is more effective when all under-represented classes are modeled jointly or when each class is modeled separately. 

(iii) \textbf{Downstream tasks:} We evaluate the effect of synthetic data on three surgical downstream applications: surgical action recognition, surgical tool presence detection, and laparoscope motion prediction.  

\paragraph{Downstream models} For surgical action (step) recognition, we employ the MViT-v2~\citep{mvitv2} model, which has shown strong performance on the SAR-RARP$50$ dataset and we report the averaged video-wise Jaccard index per class. The TAPIS model was used for the GraSP dataset, which incorporates an MViT backbone, and evaluate performance using mean average precision (mAP) averaged across videos, as described in~\citet{grasp}. For surgical tool presence detection,  the Swin Transformer (base)~\citep{swin} was opted in a multi-label classification setting, reporting the Dice score as the evaluation metric. Finally, for laparoscope motion recognition, we utilize a ResNet$3$D~\citep{resnet3d} model to classify motion categories from input clips, with mean F$1$ score as the metric. We apply inverse frequency balancing with video frame augmentation only on the real datasets during training. Especially, we add synthetic videos of under-represented to the real dataset and leave the well balanced classes undisturbed. Each model is run with three different seeds, and we report the mean and standard deviation across videos. These model choices ensure fair and robust state-of-the-art baselines for video understanding tasks. Please refer to the appendix for details on model training(~\ref{sec:diff_train}), additional experiments and evaluations(~\ref{sec:add_results}) and qualitative results (\ref{sec:qual}).

\section{Results \& Discussion}
\paragraph{Synthetic data attributes} \label{sec:syn} 
Our evaluation of different synthetic data attributes for under-represented classes in the SAR-RARP$50$ dataset is in Table~\ref{tab:syn_att}.
\begin{wraptable}{r}{0.55\linewidth}   
\caption{\small{\textbf{Attributes of synthetic} data experiment on the under-represented classes of the SAR-RARP$50$ dataset.}}
\label{tab:syn_att}
\centering
\small   
\begin{tabular}{lcccc}
\toprule
Method & A$1$ & A$5$ & A$7$ \\
\midrule
Real   & $0.32_{\pm{0.19}}$ & $0.10_{\pm{0.04}}$ & $0.32_{\pm{0.15}}$ \\
Data duplicate   & $0.32_{\pm{0.17}}$ & $0.11_{\pm{0.02}}$ & $0.32_{\pm{0.13}}$ \\
Frame shuffle   & $0.30_{\pm{0.14}}$ & $0.06_{\pm{0.09}}$ & $0.30_{\pm{0.17}}$ \\
Sparse frame   & $0.28_{\pm{0.14}}$ & $0.05_{\pm{0.05}}$ & $0.29_{\pm{0.10}}$ \\
Noisy frame   & $0.29_{\pm{0.14}}$ & $0.04_{\pm{0.05}}$ & $0.29_{\pm{0.10}}$ \\
\bottomrule
\end{tabular}
\end{wraptable}
Readers can refer to the suppl. for additional results (Sec.~\ref{ssec:syn}).
 Merely duplicating the training set does not improve performance, as it fails to introduce additional sample diversity. Frame shuffling causes a slight decline in performance, underscoring the importance of temporal consistency in video-based tasks. Similarly, injecting noise into frames or conditioning only on sparse frames results in a more substantial drop of about $3$–$5\%$. Together, these findings reveal three key aspects: (i) synthetic data must not simply replicate the training set, but rather provide \emph{data diversity}, (ii) maintaining \emph{temporal consistency} is critical, and (iii) preserving \emph{spatial structure} is essential to sustain downstream performance. Overall, this analysis underlines that downstream tasks inherently require both spatial and temporal consistency, and synthetic data must therefore satisfy both to be effective.

\paragraph{Surgical action recognition.} 
(i) \underline{\emph{SAR-RARP$50$}}:
The results of surgical action recognition task is reported in Tab.~\ref{tab:sar_tab_ori}. The SurV-Gen model with rejection sampling achieves better performance on under-represented classes compared to using synthetic samples directly, suggesting that its gains stem primarily from the sampling strategy rather than the generative model itself. 
Synthetic samples from SparseCtrl improves scores across all three underrepresented classes. Our approach, \emph{SurgiFlowVid}, even with text-only conditioning, yields performance improvements in two out of the three under-represented classes, with gains in the range of $3$–$11\%$. Adding conditional masks further enhances performance across all classes, with SurgiFlowVid conditioned on segmentation masks achieving improvements of $12\%$, $8\%$, and $10\%$ were noticed with for the under-represented classes. Performance gains are also observed in well-balanced classes, which we attribute to the mutual dependencies among actions. For instance, augmenting data for the “picking the needle” class may indirectly benefit “positioning the needle” class, as the latter can often follow in the surgical workflow. Another noteworthy observation is that modeling each class individually produces a substantial improvement in mean performance, reaching $0.53$ compared to $0.46$ with real data alone. Particularly notable is the nearly $20\%$ gain for A$7$, obtained with synthetic samples from SurgiFlowVid (RGB-frame) combined with individual-class training.
\begin{table}[tb]
\caption{\textbf{Surgical action recognition on the SAR-RARP$50$ dataset}. Under-represented classes are \hlb{highlighted}, and Jaccard index is reported. \textit{Ic} denotes individual class modeling, and RS indicates rejection sampling. Addition of synthetic samples from SurgiFlowVid indicates comprehensive gains for the under-represented classes.} 
\label{tab:sar_tab_ori}
\begin{center}
{\small
\resizebox{\linewidth}{!}{
\begin{tabular}{l cc cccccc|c}
\toprule
\multirow{2}{*}{Training data} & \multicolumn{2}{c}{Cond. type} &
\cellcolor{lightmaroon!25}\makecell{Pick\\the needle} & 
\makecell{Position\\the needle} & 
\makecell{Push\\the needle} & 
\makecell{Pull\\the needle} & 
\cellcolor{lightmaroon!25}\makecell{Cut\\the suture} & 
\cellcolor{lightmaroon!25}\makecell{Return\\the needle} & Mean. \\
\cmidrule(lr){2-3}
 & Text & Sparse mask & & & & & & &\\
\midrule
Real & -- & -- & $0.32_{\pm0.19}$ & \sbest{$0.66_{\pm0.09}$} & \sbest{$0.78_{\pm0.10}$} & $0.61_{\pm0.09}$ & $0.10_{\pm0.04}$ & $0.32_{\pm0.15}$ & $0.46_{\pm0.08}$\\ \midrule
Real $+$ Endora & -- & -- & $0.32_{\pm0.14}$ & $0.63_{\pm0.05}$ & $0.76_{\pm0.07}$ & $0.61_{\pm0.11}$ & $0.08_{\pm0.04}$ & $0.33_{\pm0.10}$ & $0.45_{\pm0.05}$\\
Real $+$ SurV-Gen (w/o RS) & \cmark & -- & $0.31_{\pm0.19}$ & $0.64_{\pm0.07}$ & $0.77_{\pm0.06}$ & $0.60_{\pm0.10}$ & $0.13_{\pm0.10}$ & $0.37_{\pm0.18}$ & $0.46_{\pm0.03}$\\
Real $+$ SurV-Gen (RS) & \cmark & -- & $0.35_{\pm0.12}$ & $0.63_{\pm0.02}$ & $0.77_{\pm0.03}$ & $0.61_{\pm0.08}$ & $0.14_{\pm0.09}$ & $0.39_{\pm0.15}$ & $0.48_{\pm0.06}$\\
Real $+$ SparseCtrl & \cmark & RGB & $0.36_{\pm0.17}$ & $0.65_{\pm0.06}$ & $0.78_{\pm0.07}$ & $0.64_{\pm0.11}$ & $0.09_{\pm0.07}$ & $0.40_{\pm0.12}$ & $0.48_{\pm0.04}$ \\
Real $+$ SparseCtrl & \cmark & Seg. & $0.36_{\pm0.14}$ & $0.61_{\pm0.12}$ & $0.77_{\pm0.07}$ & $0.63_{\pm0.11}$ & $0.16_{\pm0.11}$ & $0.38_{\pm0.17}$ & $0.49_{\pm0.04}$ \\ 
\addlinespace
Real $+$ SurgFlowVid & \cmark & -- & \sbest{$0.43_{\pm0.12}$} & $0.65_{\pm0.07}$ & $0.77_{\pm0.07}$ & $0.63_{\pm0.11}$ & $0.11_{\pm0.03}$ & $0.35_{\pm0.12}$ & $0.49_{\pm0.04}$\\
Real $+$ SurgFlowvid & \cmark & RGB & $0.36_{\pm0.17}$ & $\mathbf{0.67_{\pm0.06}}$ & \sbest{$0.78_{\pm0.08}$} & $\mathbf{0.65_{\pm0.12}}$ & $0.17_{\pm0.10}$ & \sbest{$0.42_{\pm0.12}$} & $0.51_{\pm0.04}$ \\
Real $+$ SurgFlowVid & \cmark & Seg. & $\mathbf{0.44_{\pm0.18}}$ & \sbest{$0.66_{\pm0.07}$} & $\mathbf{0.79_{\pm0.08}}$ & \sbest{$0.64_{\pm0.04}$} & \sbest{$0.18_{\pm0.09}$} & \sbest{$0.42_{\pm0.15}$} & \sbest{$0.52_{\pm0.04}$} \\ 
\addlinespace
Real $+$ SurgFlowVid (\textit{Ic}) & \cmark & -- & $0.37_{\pm0.16}$ & $0.65_{\pm0.04}$ & $0.77_{\pm0.07}$ & $0.61_{\pm0.10}$ & $0.14_{\pm0.03}$ & $0.42_{\pm0.18}$ & $0.49_{\pm0.06}$\\
Real $+$ SurgFlowvid (\textit{Ic}) & \cmark & RGB & $0.36_{\pm0.14}$ & $0.65_{\pm0.03}$ & $\mathbf{0.79_{\pm0.15}}$ & \sbest{$0.64_{\pm0.08}$} & $\mathbf{0.20_{\pm0.09}}$ & $\mathbf{0.52_{\pm0.12}}$ & $\mathbf{0.53_{\pm0.02}}$ \\
Real $+$ SurgFlowVid (\textit{Ic}) & \cmark & Seg. & $0.41_{\pm0.19}$ & $0.63_{\pm0.06}$ & $0.77_{\pm0.03}$ & $0.62_{\pm0.12}$ & $0.10_{\pm0.05}$ & $0.38_{\pm0.16}$ & $0.48_{\pm0.06}$ \\
\bottomrule
\end{tabular}
}}
\end{center}
\end{table}

\begin{table}[tb]
\caption{\textbf{Surgical step recognition on the GraSP dataset}. The best scores are in \textbf{bold} and the mAP scores are reported. Considerable performance gains are noticed for our approach with the sparse RGB frames in comparison to solely using the real dataset.}
\label{tab:grasp_act}
\begin{center}
{\small
\resizebox{\linewidth}{!}{
\begin{tabular}{l cc ccccc|c}
\toprule
\multirow{2}{*}{Training data} & \multicolumn{2}{c}{Cond. type} &
\cellcolor{lightmaroon!25}\makecell{Pull\\the suture} & 
\cellcolor{lightmaroon!25}\makecell{Tie\\the suture} & 
\cellcolor{lightmaroon!25}\makecell{Cut\\the suture} & 
\cellcolor{lightmaroon!25}\makecell{Cut\\btw. the prostate} & 
\makecell{Identify the\\ iliac artery} 
& Mean. \\
\cmidrule(lr){2-3}
 & Text & Sparse mask & & & & & & \\
\midrule
Real & -- & -- & $0.26_{\pm0.03}$ & $0.44_{\pm0.01}$ & $0.43_{\pm0.06}$ & $0.72_{\pm0.07}$ & $0.52_{\pm0.08}$ & $0.47_{\pm0.03}$\\ \midrule
Real $+$ Endora & -- & -- & $0.26_{\pm0.02}$ & $0.39_{\pm0.02}$ & $0.40_{\pm0.05}$ & $0.70_{\pm0.01}$ & $0.51_{\pm0.03}$ & $0.45_{\pm0.04}$ \\
Real $+$ SurV-Gen (w/o RS) & \cmark & -- & $0.30_{\pm0.01}$ & $0.43_{\pm0.02}$ & $0.41_{\pm0.09}$ & $0.71_{\pm0.04}$ & $0.57_{\pm0.07}$ & $0.48_{\pm0.03}$\\
Real $+$ SurV-Gen (RS) & \cmark & -- & $0.30_{\pm0.02}$ & $0.44_{\pm0.03}$ & $0.42_{\pm0.09}$ & \underline{$0.73_{\pm0.02}$} & $0.58_{\pm0.04}$ & $0.49_{\pm0.02}$\\
Real $+$ SparseCtrl & \cmark & RGB & $0.27_{\pm0.01}$ & $0.43_{\pm0.01}$ & $0.40_{\pm0.09}$ & $0.71_{\pm0.04}$ & $0.55_{\pm0.04}$ & $0.46_{\pm0.04}$ \\
\addlinespace
Real $+$ SurgFlowVid & \cmark & -- & $0.30_{\pm0.01}$ & $0.43_{\pm0.03}$ & $\underline{0.44_{\pm0.09}}$ & $0.69_{\pm0.04}$ & $\underline{0.60_{\pm0.07}}$ & $0.49_{\pm0.04}$\\
Real $+$ SurgFlowvid & \cmark & RGB & $\mathbf{0.33_{\pm0.01}}$ & $\mathbf{0.48_{\pm0.02}}$ & $\mathbf{0.47_{\pm0.01}}$ & $\mathbf{0.74_{\pm0.02}}$ & $\underline{0.60_{\pm0.05}}$ & $\mathbf{0.52_{\pm0.04}}$ \\
\addlinespace
Real $+$ SurgFlowVid (\textit{Ic}) & \cmark & -- & $\underline{0.31_{\pm0.04}}$ & $0.41_{\pm0.03}$ & $0.42_{\pm0.04}$ & $0.72_{\pm0.04}$ & $\mathbf{0.61_{\pm0.05}}$ & $0.49_{\pm0.03}$\\
Real $+$ SurgFlowvid (\textit{Ic}) & \cmark & RGB & $\underline{0.31_{\pm0.01}}$ & $\underline{0.45_{\pm0.01}}$ & $0.43_{\pm0.03}$ & $0.72_{\pm0.02}$ & $0.55_{\pm0.05}$ & $\underline{0.50_{\pm0.02}}$ \\
\bottomrule
\end{tabular}
}}
\end{center}
\end{table}

\begin{figure}[!htbp]
    \centering
    \includegraphics[width=\textwidth, keepaspectratio]{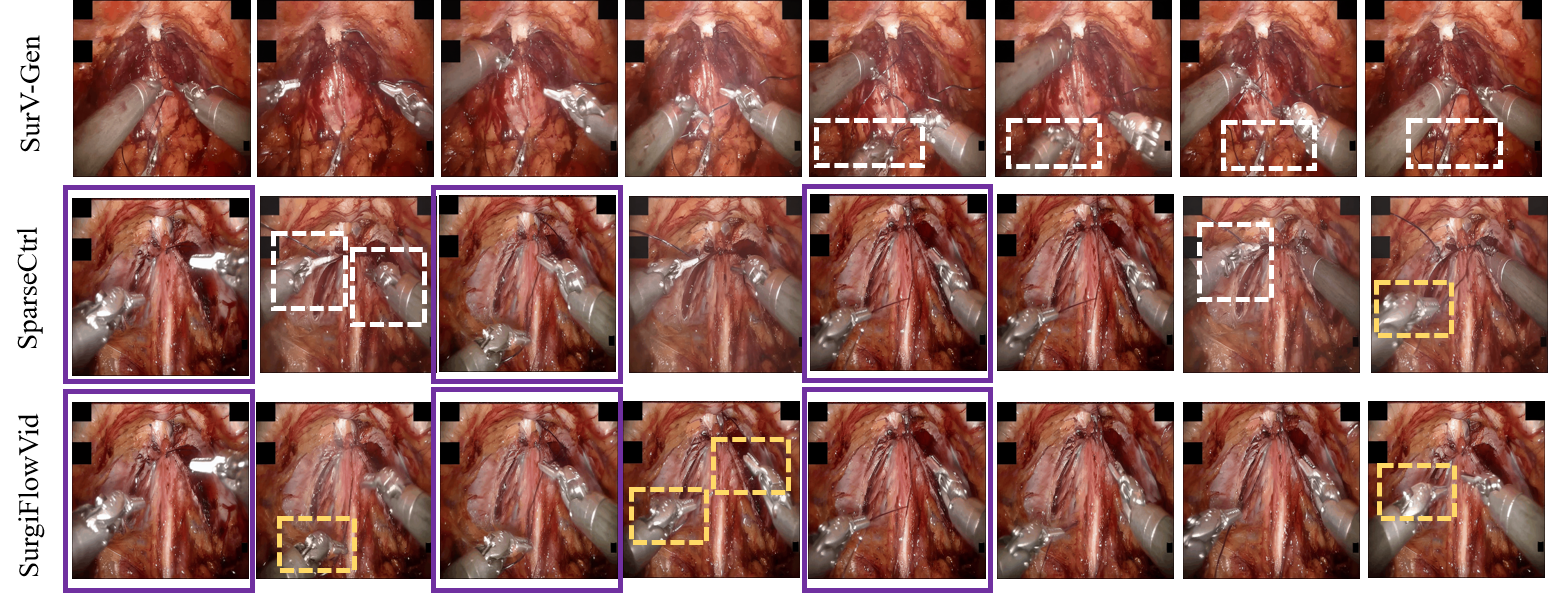}
    \caption{\textbf{Qualitative results} of the action ``tie the suture.'' Purple boxes denote the sparse RGB conditioning frames. Suprious tools are generated in SurV-Gen (white box, row $1$), while SparseCtrl alters tool types compared to the conditioning frames (white box, row $2$), reflecting limited spatial consistency. SurgiFlowVid preserves both spatial and temporal structure, with consistent tools maintained across generated frames (yellow boxes, row $3$). }
    \label{fig:qual_comp}
\end{figure}

(ii) \underline{\emph{GraSP}}: The effect of adding synthetic samples on the GraSP dataset is shown in Tab.~\ref{tab:grasp_act} and the qualitative results are shown in Fig.~\ref{fig:qual_comp}. Incorporating samples from SurV-Gen yields small performance gains, whereas adding data generated by Endora (the unconditional baseline) or SparseCtrl with RGB-frame conditioning results in a decline in mAP score. 
By contrast, our method, SurgiFlowVid, achieves improvements in two of the four underrepresented classes even with text conditioning. Furthermore, with sparse segmentation masks SurgiFlowVid achieves performance gains across all under-represented classes. These results highlight the combined value of our dual-prediction and spar encoder modules, which enables the model to learn spatio-temporal relationships from limited data more effectively.


\paragraph{Surgical tool presence detection}
\begin{table}[!htbp]
\caption{\textbf{Surgical tool presence detection on SAR-RARP$50$ dataset}. Our approach with seg. conditioning outperforms the baseline across all seven tool categories.}
\label{tab:tool_sar}
\begin{center}
{\small
\resizebox{\linewidth}{!}{
\begin{tabular}{cccccccccc|c}
\toprule
Training data &
\makecell{Tool\\clasper} & 
\makecell{Tool\\wrist} & 
\makecell{Tool\\shaft} & 
\makecell{Suturing\\needle} & 
\makecell{Thread} & 
\cellcolor{blue!10}\makecell{Suction\\tool} & \cellcolor{blue!10}\makecell{Needle\\holder} & \cellcolor{blue!10}Clamps & \cellcolor{blue!10}Catheter & Mean \\
\midrule
Real & $0.85_{\pm0.10}$ & $0.84_{\pm0.09}$ & $0.88_{\pm0.07}$ & $0.70_{\pm0.15}$ & $0.75_{\pm0.12}$ & $0.69_{\pm0.11}$ & $0.66_{\pm0.07}$ & $0.44_{\pm0.11}$ & $0.46_{\pm0.08}$ & $0.69_{\pm0.06}$\\ \midrule
\makecell{Real \\$+$ SparseCtrl(Seg)}  & $0.87_{\pm0.11}$ & $0.83_{\pm0.05}$ & $\mathbf{0.89_{\pm0.06}}$ & $0.73_{\pm0.12}$ & $0.80_{\pm0.13}$ & $\mathbf{0.79_{\pm0.10}}$ & $0.74_{\pm0.09}$ & $0.69_{\pm0.08}$ & $0.50_{\pm0.12}$ & $0.74_{\pm0.03}$\\ 
\makecell{Real \\$+$ SurgFlowVid(Seg)} & $\mathbf{0.88_{\pm0.09}}$ & $\mathbf{0.85_{\pm0.07}}$ & $0.88_{\pm0.10}$ & $\mathbf{0.75_{\pm0.11}}$ & $\mathbf{0.81_{\pm0.09}}$ & $0.78_{\pm0.15}$ & $\mathbf{0.75_{\pm0.04}}$ & $\mathbf{0.73_{\pm0.10}}$ & $\mathbf{0.59_{\pm0.05}}$ & $\mathbf{0.79_{\pm0.04}}$\\
\bottomrule
\end{tabular}
}}
\end{center}
\end{table}

\begin{table}[!htbp]
\caption{\textbf{Surgical tool presence detection on GraSP dataset}. Combining synthetic data from SurgiFlowVid yields marked improvements in dice scores.}
\label{tab:tool_grasp}
\begin{center}
{\small
\resizebox{\linewidth}{!}{
\begin{tabular}{cccccccc|c}
\toprule
Training data &
\makecell{Bipolar\\forceps} & 
 \makecell{L.needle\\driver} & 
\makecell{Mono\\curved scissors} & 
\cellcolor{blue!10}\makecell{Prograsp\\forceps} &
\cellcolor{blue!10}\makecell{Suction\\inst.} & 
\cellcolor{blue!10}\makecell{Clip\\applier} & \cellcolor{blue!10}\makecell{Laparoscopic\\inst.} &  Mean \\
\midrule
Real & $0.94_{\pm0.01}$ & $0.56_{\pm0.03}$ & $0.95_{\pm0.02}$ & $0.72_{\pm0.02}$ & $0.71_{\pm0.03}$ & $0.34_{\pm0.09}$ & $0.56_{\pm0.04}$ & $0.68_{\pm0.10}$\\ 
\makecell{Real \\$+$ SparseCtrl(Seg)}  & $\mathbf{0.95_{\pm0.02}}$ & $0.56_{\pm0.02}$ & $0.97_{\pm0.01}$ & $0.75_{\pm0.03}$ & $\mathbf{0.74_{\pm0.07}}$ & $0.35_{\pm0.02}$ & $\mathbf{0.60_{\pm0.05}}$ & $0.70_{\pm0.04}$ \\ 
\makecell{Real \\$+$ SurgFlowVid(Seg)} & $0.94_{\pm0.01}$ & $\mathbf{0.58_{\pm0.02}}$ & $\mathbf{0.98_{\pm0.01}}$ & $\mathbf{0.78_{\pm0.01}}$ & $0.73_{\pm0.04}$ & $\mathbf{0.37_{\pm0.03}}$ & $\mathbf{0.60_{\pm0.02}}$ & $\mathbf{0.72_{\pm0.02}}$\\
\bottomrule
\end{tabular}
}}
\end{center}
\end{table}
The results of the surgical tool presence detection task on the GraSP and SAR-RARP$50$ datasets are shown in Tab.~\ref{tab:tool_sar} and Tab.\ref{tab:tool_grasp}, respectively. Overall, the addition of synthetic samples from generative models leads to consistent performance improvements. This trend can be explained by the fact that the generated surgical videos naturally increase the occurrence of individual tools within the training set. On SAR-RARP$50$, our approach, SurgiFlowVid, achieves a $10$-point improvement over using only real data, compared to a $5$-point gain from SparseCtrl. Notably, SparseCtrl’s reliance on sparse conditioning yields only limited benefits, improving performance for a single under-represented class out of four. These findings further underscore the importance of generating videos with coherent spatio-temporal context for downstream tool detection models to perform effectively. For the GraSP dataset, improvements with synthetic samples are more subtle. SparseCtrl yields modest gains, while SurgiFlowVid achieves a $6\%$ improvement for the prograsp forceps class and an overall $4\%$ improvement across the dataset. \textit{Together, these results highlight that SurgiFlowVid not only improves rare-class detection but also strengthens overall tool recognition performance.}

\begin{figure}[t]
  \centering
  \begin{minipage}[t]{0.50\textwidth}
    \centering
    \includegraphics[width=\linewidth, keepaspectratio]{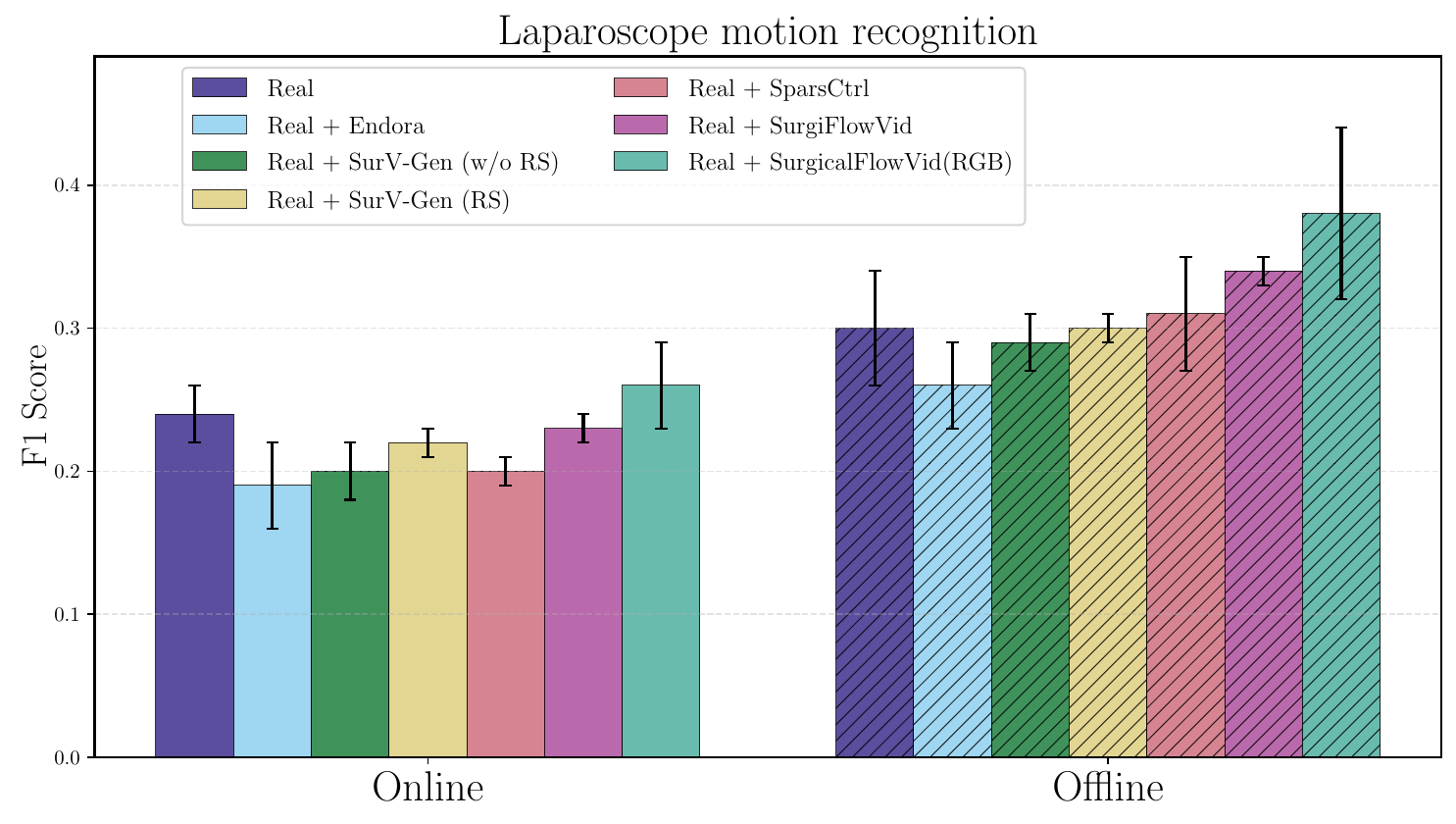}
    \caption{\small{\textbf{Laparoscope motion prediction} in \emph{online} (left) and \emph{offline} (right) fashion on the AutoLapro dataset.}}
    \label{fig:auto_comp}
  \end{minipage}
  \hfill
  \begin{minipage}[t]{0.45\textwidth}
    \centering
    \includegraphics[width=\linewidth, keepaspectratio]{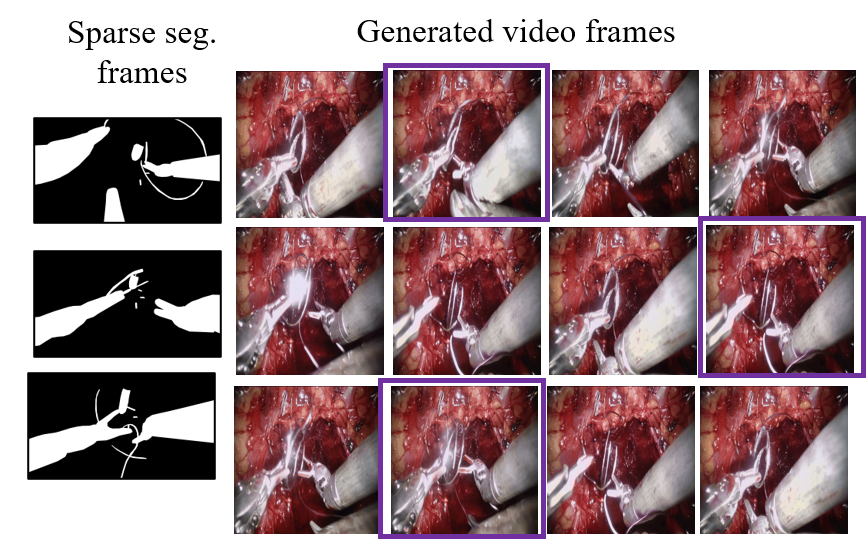}
    \caption{\small{Tool types match the sparse seg. frames, but their position shifts, causing a failure case.}}
    \label{fig:failcase}
  \end{minipage}
\end{figure}

\paragraph{Laparoscope motion prediction}
Fig.~\ref{fig:auto_comp} presents the results of laparoscope motion detection on the AutoLaparo dataset. Among the baselines, SurV-Gen (RS) achieves better performance than Endora, while SparseCtrl with RGB-frame conditioning performs best on the online recognition task. 
Our approach, SurgiFlowVid, already outperforms SurV-Gen with text-only conditioning, and the RGB-mask conditioned version surpasses all baselines. 
Similar trends are observed for the offline recognition task, where both F$1$ scores are higher compared to the online setting. This suggests that providing a longer temporal context enables the downstream model to classify laparoscope motion more accurately. Overall, these findings demonstrate that SurgiFlowVid can effectively adapt to smaller datasets while offering substantial benefits for developing automatic field-of-view control systems. \textit{This highlights the practical utility of our method in developing real-time surgical assistance systems.}

\paragraph{Limitations} While our approach demonstrates performance gains for underrepresented classes, it also has certain limitations. Currently, we generate only short video clips of about four seconds. Extending it with autoregressive generation could enable longer sequences, that are important for tasks such as surgical phase recognition. Moreover, the sparsity of segmentation frames lead to incorrect tool position generation (see Fig.~\ref{fig:failcase}), which could be mitigated by richer conditional signals such as tool kinematics or feature-level injections—directions we leave for future investigation.
\section{Conclusion}
In this work, we addressed the critical challenge of data imbalance in surgical datasets by generating synthetic video samples of under-represented classes with our proposed framework, \emph{SurgiFlowVid}. The framework generates spatially and temporally coherent videos through a dual-prediction diffusion U-Net that jointly models RGB frames and optical flow, while a sparse visual encoder enables controllable generation using only the limited conditional signals typically available in surgical datasets. Extensive experiments across three datasets and downstream tasks—surgical action recognition, tool presence detection, and laparoscope motion prediction—demonstrate consistent improvements over strong baselines. By bridging advances in machine learning with the needs of surgical data science, this work helps address the scarcity of data on rare events and moves toward more robust surgical video understanding models.

\section{Reproducibility Statement}
All the information such as models, hyper-parameters and datasets needed to reproduce this work has been included in the appendix. 

\bibliographystyle{abbrvnat}
\bibliography{ref}

\clearpage
\appendix

The appendix is structured as follows:
\begin{enumerate}[label=\Alph*.]
    \item Extended related work section \dotfill \ref{sec:add_related_work}
    \item Additional results
    \begin{enumerate}[label=\arabic*.]
        \item Synthetic data attributes \dotfill \ref{ssec:syn}
        \item Gynsurg – action recognition dataset \dotfill \ref{sec:add_data}
        \item Different downstream model architecture \dotfill \ref{sec:model_Arch}
        \item Video metrics \dotfill \ref{sec:add_metrics}
        \item Sparse frame ablation \dotfill \ref{sec:sparse_frame}
        \item Model analysis \dotfill \ref{sec:model_analysis}
        \item Image quality results \dotfill \ref{sec:image_quality}
        \item Lap. motion detection \dotfill \ref{sec:lap_motion}
    \end{enumerate}
    \item Dataset information \dotfill \ref{sec:dataset_add}
    \item Model training details
    \begin{enumerate}[label=\arabic*.]
        \item Diffusion image training \dotfill \ref{sec:diffusion_image}
        \item Diffusion video pre-training \dotfill \ref{sec:diffusion_pretrainin}
        \item SurgiFlowVid training \dotfill \ref{sec:surgiflow_train}
        \item Downstream model \dotfill \ref{sec:downstream}
    \end{enumerate}
    \item Qualitative results \dotfill \ref{sec:qual}
\end{enumerate}

\section{Extended Related Work} \label{sec:add_related_work}
\paragraph{Video diffusion models} Diffusion-based video generation methods have recently demonstrated strong efficiency and scalability by operating in continuous latent spaces~\citep{ho2020denoising,sd}. Early work by~\citep{ho2022video} extended pixel-space diffusion to videos using probabilistic DMs, while~\citep{harvey2022flexible} proposed generating sparse frames with interpolation, though limited to low-resolution synthetic datasets. Large-scale efforts such as Make-A-Video~\citep{singer2022make} and Imagen Video~\citep{ho2022imagen} employ cascaded super-resolution pipelines built on DALLE-2~\citep{ramesh2022hierarchical} and Imagen~\citep{saharia2022photorealistic}, respectively, but require billions of parameters and massive compute resources. Stable Video Diffusion~\citep{blattmann2023stable} has been widely adopted in the natural image/video community, while several closed-source systems—such as MovieGen~\citep{moviegen}, Pika~\citep{pika}, and Gen (Runway)~\citep{runway}, Veo~\citep{veo}—achieve high-quality generation conditioned on diverse modalities ranging from text to depth maps.

On the open-source side, AnimateDiff~\citep{guo2023animatediff} and SparseCtrl~\citep{sparse} extend image diffusion models to videos, while OpenSora~\citep{zheng2024open} represents a large community-driven effort to replicate Sora~\citep{sora}. The CogVideo family~\citep{yang2024cogvideox,hong2022cogvideo} introduces expert transformer architectures for video synthesis and has been adopted in prior surgical applications~\citep{hierasurg,iliash2024interactive}. However, CogVideo is a $5$B parameter model requiring vast datasets and heavy compute, making it impractical for limited surgical data where overfitting is a risk. We inspire our approach from the more recently proposed methods such as FlowVid~\citep{liang2024flowvid} and VideoJam~\citep{chefer2025videojam}. FlowVid proposed a flow warped video-to-video generation framework, wherein optical flow was used to maintain the structure of objects between frames during translation. This framework trained on a corpus of $10$M videos. The primary application of this work differs from ours such that we intend to generate new videos with conditional signals. Secondly, VideoJam explored video prediction with a DiT-based architecture~\citep{peebles2023scalable}, but its $30$B parameter model was trained on $100$M videos, produces only $256 \times 256$ outputs, and lacks controllability—an essential requirement for surgical applications.

In contrast, our work targets the surgical domain under constrained compute budgets, focusing on the critical issue of data imbalance. We build upon small-scale surgical video diffusion models and introduce a sparse, controllable framework tailored to generate under-represented surgical classes. To the best of our knowledge, we are the first to introduce a conditional video diffusion framework to mitigate the data imbalance issue for surgical application. While future work could explore scaling to larger models, our approach demonstrates a practical pathway toward improving surgical video understanding in realistic healthcare settings.

\paragraph{Data imbalance} The presence of rare classes is a common challenge in real-world datasets. In classification, oversampling is frequently used to mitigate this issue by sampling under-represented classes more often during training~\citep{oversampling}. Standard augmentation methods such as horizontal flipping, random resizing, and cropping are widely used, while regional dropout methods~\citep{zhong2020random} randomly remove image regions to improve robustness and generalization. More advanced strategies, including RandAugment~\citep{randaugment} and AutoAugment~\citep{autoaugment}, apply diverse pixel-level operations (e.g., rotation, shear, translation, color jitter) through either random selection or learned policies. Other approaches combine multiple images, such as Mixup~\citep{mixup}, which blends both pixel values and labels, and CutMix~\citep{cutmix}, which replaces patches from one image with regions from another, maximizing pixel efficiency while mixing labels. These augmentation strategies have been specifically proposed for image classification tasks. Readers can refer to~\citep{chen2024survey} for a detailed survey.

Within the surgical domain, class imbalance is particularly prevalent due to challenges in data collection (e.g., reliance on single-center data), the rarity of specific surgical events, and ethical or legal restrictions on data sharing~\citep{handling,maier2017surgical}. Such imbalance often degrades the performance of downstream models. While augmentation and re-sampling strategies have been shown to improve medical imaging tasks~\citep{handling}, surgical video understanding tasks lacks dedicated augmentation approaches. Prior attempts have used synthetic data, for instance via image-to-image translation, to complement real datasets for only surgical instrument segmentation tasks~\citep{ssis, robotic, rethinking}. In this work, we establish a strong baseline for real datasets by combining curated image-level augmentation techniques with inverse frequency balancing, which up-weights under-represented classes. We use this strategy only during the training of real datasets. To directly assess the utility of synthetic data as a complementary augmentation strategy, we merge generated videos with real data without applying further augmentations.

\section{Additional results} \label{sec:add_results}
\subsection{Synthetic data attributes}\label{ssec:syn} The results on different aspects of synthetic data for the SAR-RARP$50$ dataset are presented in Tab.~\ref{tab:sar_tab_att}. Performance remains unchanged when the training data is merely duplicated, a trend consistent across most classes. In contrast, perturbations to either the spatial or temporal structure of the videos result in clear performance degradation. This behavior aligns with the role of the downstream model, which relies on both spatial structure (e.g., the arrangement of organs) and temporal dynamics (e.g., tissue motion and single or multi-tool interactions) to classify an action. Notably, the action class ``cutting the suture,'' which is already highly imbalanced, suffers a substantial drop in performance when frame-level noise is introduced. Similar results were noticed for the GraSP dataset (Tab.~\ref{tab:grasp_tab_app}). Interestingly we noticed for shuffling the frames lead to a small improvement in scores for two of the under-represented classes. This results could also be attributed to the downstream model architecture difference between the TAPIS model and the plain MViT model. However, overall these findings highlight that synthetic data cannot simply replicate training samples, nor can it exhibit spatial or temporal inconsistencies, if it is to provide meaningful benefits for downstream tasks. 

\begin{table}[!htbp]
\caption{\textbf{Attributes of synthetic} data experiment on the SAR-RARP$50$ dataset. Merely replicating the training data does not lead to any improvement in performance. The degradation of the spatial or temporal structure leads to a decline in downstream model performance.}
\label{tab:sar_tab_att}
\begin{center}
{\small
\resizebox{\linewidth}{!}{
\begin{tabular}{lcccccc|c}
\toprule
\multirow{2}{*}{Training data} &
\cellcolor{lightmaroon!20}\makecell{Pick\\the needle} & 
\makecell{Position\\the needle} & 
\makecell{Push\\the needle} & 
\makecell{Pull\\the needle} & 
\cellcolor{lightmaroon!20}\makecell{Cut\\the suture} & 
\cellcolor{lightmaroon!20}\makecell{Return\\the needle} & Mean. \\
\midrule
Real & $0.32_{\pm0.19}$ & $0.66_{\pm0.09}$ & $0.78_{\pm0.10}$ & $0.61_{\pm0.09}$ & $0.10_{\pm0.04}$ & $0.32_{\pm0.15}$ & $0.46_{\pm0.08}$\\ \midrule
Data duplication & $0.32_{\pm0.17}$ & $0.60_{\pm0.03}$ & $0.78_{\pm0.08}$ & $0.61_{\pm0.10}$ & $0.10_{\pm0.03}$ & $0.31_{\pm0.11}$ & $0.45_{\pm0.06}$\\
Frame shuffle  & $0.30_{\pm0.19}$ & $0.63_{\pm0.08}$ & $0.74_{\pm0.11}$ & $0.60_{\pm0.08}$ & $0.06_{\pm0.09}$ & $0.30_{\pm0.17}$ & $0.43_{\pm0.04}$\\
Sparse frame  & $0.28_{\pm0.14}$ & $0.60_{\pm0.07}$ & $0.70_{\pm0.04}$ & $0.59_{\pm0.09}$ & $0.05_{\pm0.05}$ & $0.29_{\pm0.10}$ & $0.42_{\pm0.03}$\\
Noisy frame & $0.29_{\pm0.14}$ & $0.62_{\pm0.07}$ & $0.76_{\pm0.04}$ & $0.60_{\pm0.09}$ & $0.04_{\pm0.05}$ & $0.29_{\pm0.10}$ & $0.43_{\pm0.02}$ \\
\bottomrule
\end{tabular}
}}
\end{center}
\end{table}

\begin{table}[!htbp]
\caption{\textbf{Attributes of synthetic} data experiment on the GraSP dataset.}
\label{tab:grasp_tab_app}
\begin{center}
{\small
\resizebox{\linewidth}{!}{
\begin{tabular}{lccccc|c}
\toprule
\multirow{2}{*}{Training data} &
\cellcolor{lightmaroon!20}\makecell{Pull\\the suture} & 
\cellcolor{lightmaroon!20}\makecell{Tie\\the suture} & 
\cellcolor{lightmaroon!20}\makecell{Cut\\the suture} & 
\cellcolor{lightmaroon!20}\makecell{Cut\\btw.the prostate} & 
\makecell{Identify\\iliac artery} & Mean. \\
\midrule
Real & $0.26_{\pm0.03}$ & $0.44_{\pm0.01}$ & $0.43_{\pm0.06}$ & $0.72_{\pm0.07}$ & $0.52_{\pm0.08}$ & $0.46_{\pm0.08}$\\ \midrule
Data duplication & $0.25_{\pm0.02}$ & $0.44_{\pm0.02}$ & $0.43_{\pm0.05}$ & $0.71_{\pm0.06}$ & $0.52_{\pm0.04}$ & $0.46_{\pm0.04}$\\
Frame shuffle  & $0.27_{\pm0.04}$ & $0.40_{\pm0.02}$ & $0.42_{\pm0.01}$ & $0.69_{\pm0.03}$ & $0.53_{\pm0.04}$ & $0.46_{\pm0.02}$\\
Sparse frame  & $0.24_{\pm0.02}$ & $0.38_{\pm0.03}$ & $0.40_{\pm0.02}$ & $0.68_{\pm0.02}$ & $0.48_{\pm0.01}$  & $0.43_{\pm0.02}$\\
Noisy frame & $0.20_{\pm0.04}$ & $0.35_{\pm0.05}$ & $0.34_{\pm0.06}$ & $0.66_{\pm0.03}$ & $0.46_{\pm0.05}$ & $0.40_{\pm0.04}$ \\
\bottomrule
\end{tabular}
}}
\end{center}
\end{table}

\subsection{Additional Surgical action dataset} \label{sec:add_data} We further evaluated surgical action recognition on the GynSurg dataset~\citep{gynsurg}, which consists of laparoscopic gynecological procedures with four annotated actions: coagulation (P$1$), needle passing (P$2$), suction/irrigation (P$3$), and transection (P$4$). The classes P$3$ and P$4$ are under-represented. Each action is provided as short $3$-second video clips, making the dataset well-suited for action recognition. Importantly, this dataset differs substantially from SAR-RARP$50$ and GraSP in terms of anatomy, environment, tool usage, and camera motion, allowing us to demonstrate the generalizability of our approach across diverse surgical settings. We adopt the MViTv$2$ model as the downstream architecture.

\begin{figure}
    \centering
    \includegraphics[width=\textwidth, keepaspectratio]{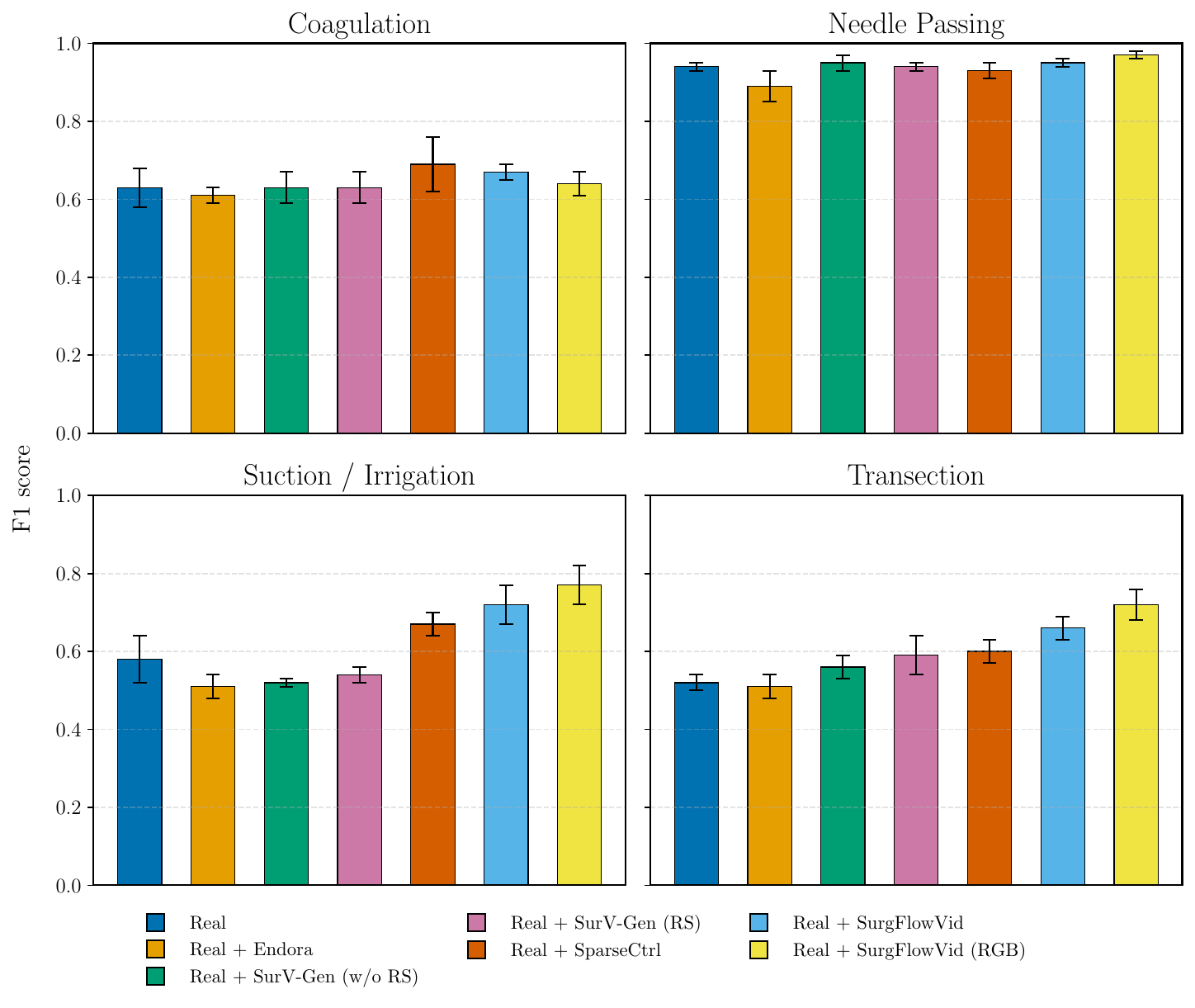}
    \caption{\textbf{Surgical action recognition} results on the GynSurg dataset, reported using the F$1$ score. The under-represented classes are ``Suction'' and ``Transection''. The addition of synthetic samples for both the balanced classes shows smaller improvements. However, the synthetic video samples from our approach (SurgiFlowVid) with text conditioning improves performance for both under-represented classes, while sparse RGB frame conditioning yields gains of up to $20$ points in comparison to using only the real dataset.}
    \label{fig:gyn_act}
\end{figure}

Results are reported in Fig.~\ref{fig:gyn_act}. Synthetic samples from SparseCtrl improve performance by $8$–$9\%$ for the under-represented classes. In contrast, our method with text conditioning achieves consistent gains across all four classes, raising the average Jaccard score to $0.72$ compared to $0.66$ with real data only. Conditioning with RGB frames yields further improvements of nearly $20$ points for P$3$ and P$4$. These results highlight the advantage of combining dual-prediction with sparse visual encoding to generate synthetic videos that preserve both spatial and temporal consistency.

\subsection{Model architecture} \label{sec:model_Arch}
We further analyzed the impact of synthetic data using a different architecture for action recognition on SAR-RARP$50$. Since the MViT model is purely transformer-based, we tested whether synthetic samples introduce any architectural bias by comparing against X3D~\citep{x3d}, a lightweight $3$D convolutional model with only $3$M parameters (vs. $30$M for MViT). The evaluation setup remained identical to previous experiments. The results are shown in Tab.~\ref{tab:sar_tab_x3d}. Compared to Tab.~\ref{tab:sar_tab_ori}, the mean Jaccard score with real data dropped to $0.38$ for X3D (vs. $0.46$ for MViT), as expected given the smaller capacity of X3D.

\begin{table}[!htbp]
\caption{\textbf{Influence of model architecture}. The surgical action recognition task on the SAR-RARP$50$ dataset using X$3$D model. The Jaccard index is reported. Best and second-best scores are highlighted in \hlblue{blue} and \hlgreen{green}, respectively. Under-represented classes are indicated with \hl{shade}. We notice similar trends to Tab.~\ref{tab:sar_tab_ori}, where the addition of samples from our approach leads to performance gains for all the under-represented classes.} 
\label{tab:sar_tab_x3d}
\begin{center}
{\small
\resizebox{\linewidth}{!}{
\begin{tabular}{l cc cccccc|c}
\toprule
\multirow{2}{*}{Training data} & \multicolumn{2}{c}{Cond. type} &
\cellcolor{lightmaroon!20}\makecell{Pick\\the needle} & 
\makecell{Position\\the needle} & 
\makecell{Push\\the needle} & 
\makecell{Pull\\the needle} & 
\cellcolor{lightmaroon!20}\makecell{Cut\\the suture} & 
\cellcolor{lightmaroon!20}\makecell{Return\\the needle} & Mean. \\
\cmidrule(lr){2-3}
 & Text & Sparse mask & & & & & & &\\
\midrule
Real & -- & -- & $0.22_{\pm0.01}$ & $0.54_{\pm0.08}$ & $0.75_{\pm0.07}$ & $0.51_{\pm0.13}$ & $0.10_{\pm0.02}$ & $0.20_{\pm0.12}$ & $0.38_{\pm0.06}$\\ \midrule
Real $+$ Endora & -- & -- & $0.19_{\pm0.04}$ & $0.53_{\pm0.02}$ & $0.75_{\pm0.05}$ & $0.50_{\pm0.10}$ & $0.09_{\pm0.05}$ & $0.18_{\pm0.04}$ & $0.38_{\pm0.06}$\\
Real $+$ SurV-Gen (w/o RS) & \cmark & -- & $0.22_{\pm0.10}$ & $0.54_{\pm0.04}$ & $0.75_{\pm0.02}$ & $0.51_{\pm0.08}$ & $0.11_{\pm0.09}$ & $0.19_{\pm0.08}$ & $0.39_{\pm0.07}$\\
Real $+$ SurV-Gen (RS) & \cmark & -- & $0.23_{\pm0.11}$ & $0.54_{\pm0.06}$ & $0.74_{\pm0.07}$ & $0.52_{\pm0.11}$ & $0.10_{\pm0.09}$ & $0.23_{\pm0.16}$ & $0.39_{\pm0.06}$\\
Real $+$ SparseCtrl & \cmark & RGB & \ssbest{$0.34_{\pm0.17}$} & $0.60_{\pm0.07}$ & $0.77_{\pm0.08}$ & $0.58_{\pm0.09}$ & $0.08_{\pm0.05}$ & $0.23_{\pm0.16}$ & $0.43_{\pm0.03}$ \\
Real $+$ SparseCtrl & \cmark & Seg. & $0.33_{\pm0.14}$ & $0.58_{\pm0.06}$ & $0.75_{\pm0.07}$ & $0.57_{\pm0.13}$ & $0.09_{\pm0.03}$ & $0.28_{\pm0.17}$ & $0.43_{\pm0.04}$\\ \midrule
Real $+$ SurgFlowVid & \cmark & -- & \ssbest{$0.34_{\pm0.13}$} & $0.58_{\pm0.06}$ & $0.75_{\pm0.05}$ & $0.55_{\pm0.13}$ & \best{$0.18_{\pm0.09}$} & \ssbest{$0.29_{\pm0.12}$} & \ssbest{$0.45_{\pm0.04}$} \\
Real $+$ SurgFlowvid & \cmark & RGB & $0.30_{\pm0.19}$ & $0.58_{\pm0.06}$ & $0.74_{\pm0.07}$ & $0.58_{\pm0.10}$ & $0.10_{\pm0.08}$ & $0.26_{\pm0.17}$ & $0.43_{\pm0.02}$ \\
Real $+$ SurgFlowVid & \cmark & Seg. & \best{$0.39_{\pm0.12}$} & $0.60_{\pm0.05}$ & $0.76_{\pm0.08}$ & $0.56_{\pm0.12}$ & \ssbest{$0.13_{\pm0.05}$} & \best{$0.35_{\pm0.12}$} & \best{$0.47_{\pm0.02}$} \\ 
\bottomrule
\end{tabular}
}}
\end{center}
\end{table}

Synthetic data from SparseCtrl led to modest improvements, while SurgiFlowVid with text conditioning provided only subtle gains. However, consistent with trends in Tab.~\ref{tab:sar_tab_ori}, adding sparse RGB or segmentation masks as conditional signals in SurgiFlowVid yielded considerable improvements across the under-represented classes. Similar trends were noticed when we performed individual class modelling with the results shown in Tab.~\ref{tab:sar_tab_x3d_ic}. These findings suggest that performance gains from synthetic data are not biased toward a specific architecture; instead, both transformer- and convolution-based models benefit from the spatial and temporal consistency encoded in synthetic videos. For the GraSP dataset, we opted to use the TAPIS model as proposed in \citep{grasp} as this model performed in par with other convolutional architectues.

Using the features extracted from downstream models, temporal models are trained to enhance action recognition further. However, the reported performance improvements were minimal~\citep{funke2025tunes}, and we therefore did not pursue such experiments in this study. Future work could explore this direction in greater depth, focusing on identifying which features from synthetic data are most beneficial for improving the generation process. Additionally, incorporating temporal learning strategies on top of these features may provide further gains for surgical action recognition tasks.

\begin{table}[!htbp]
\caption{\textbf{Influence of model architecture}. The surgical action recognition task on the SAR-RARP$50$ dataset using X$3$D model with \emph{individual class modelling}. The Jaccard index is reported. We notice smaller gains for the action ``cut the suture'' (see Tab.~\ref{tab:sar_tab_x3d}) by modeling each of the under-represented classes separately.} 
\label{tab:sar_tab_x3d_ic}
\begin{center}
{\small
\resizebox{\linewidth}{!}{
\begin{tabular}{l cc cccccc|c}
\toprule
\multirow{2}{*}{Training data} & \multicolumn{2}{c}{Cond. type} &
\cellcolor{lightmaroon!20}\makecell{Pick\\the needle} & 
\makecell{Position\\the needle} & 
\makecell{Push\\the needle} & 
\makecell{Pull\\the needle} & 
\cellcolor{lightmaroon!20}\makecell{Cut\\the suture} & 
\cellcolor{lightmaroon!20}\makecell{Return\\the needle} & Mean. \\
\cmidrule(lr){2-3}
 & Text & Sparse mask & & & & & & &\\
\midrule
Real & -- & -- & $0.22_{\pm0.01}$ & $0.54_{\pm0.08}$ & $0.75_{\pm0.07}$ & $0.51_{\pm0.13}$ & $0.10_{\pm0.02}$ & $0.20_{\pm0.12}$ & $0.38_{\pm0.06}$\\ \midrule
Real $+$ SurV-Gen (RS) & \cmark & -- & $0.25_{\pm0.12}$ & $0.54_{\pm0.03}$ & $0.76_{\pm0.09}$ & $0.51_{\pm0.09}$ & $0.10_{\pm0.13}$ & $0.24_{\pm0.18}$ & $0.40_{\pm0.05}$\\
Real $+$ SparseCtrl & \cmark & RGB & $0.30_{\pm0.16}$ & $0.59_{\pm0.07}$ & $0.75_{\pm0.06}$ & $0.57_{\pm0.11}$ & $0.10_{\pm0.09}$ & $0.21_{\pm0.12}$ & $0.42_{\pm0.03}$ \\
Real $+$ SparseCtrl & \cmark & Seg. & $0.30_{\pm0.17}$ & $0.57_{\pm0.04}$ & $0.76_{\pm0.07}$ & $0.57_{\pm0.09}$ & \ssbest{$0.20_{\pm0.05}$} & \ssbest{$0.37_{\pm0.10}$} & $0.46_{\pm0.01}$\\ \midrule
Real $+$ SurgFlowvid & \cmark & RGB & \best{$0.40_{\pm0.16}$} & $0.56_{\pm0.02}$ & $0.75_{\pm0.04}$ & $0.56_{\pm0.16}$ & \best{$0.23_{\pm0.13}$} & $0.35_{\pm0.15}$ & \best{$0.48_{\pm0.02}$} \\
Real $+$ SurgFlowVid & \cmark & Seg. & \ssbest{$0.39_{\pm0.11}$} & $0.59_{\pm0.04}$ & $0.77_{\pm0.03}$ & $0.55_{\pm0.10}$ & $0.15_{\pm0.06}$ & \best{$0.40_{\pm0.10}$} & \ssbest{$0.48_{\pm0.05}$} \\ 
\bottomrule
\end{tabular}
}}
\end{center}
\end{table}

\subsection{Video metrics} \label{sec:add_metrics}
We assess the temporal performance of the model using  \textit{Segmental F1@K} score. This metric penalizes
both out-of-order predictions and over-segmentation. Segmental F1@K quantifies the temporal overlap between predicted and ground-truth segments, while being less sensitive to small boundary shifts caused by
annotation noise. The metric is defined as, 
\begin{equation}
\text{SegmentalF1@K} = 
\frac{2 \times (\text{Pr} \times \text{Rc})}
     {(\text{Pr} + \text{Rc})},
\end{equation}
where Pr and Rc denotes precision and recall. A prediction is considered a true positive (TP) if the IoU exceeds the threshold \(T = K/100\); otherwise, it is counted as a false positive (FP). The results of the recognition task are shown in Tab.\ref{tab:sar_tab_segf1} and Tab.\ref{tab:sar_tab_segf1_seg}. 
Compared to using only the real dataset, the addition of synthetic samples leads to smaller improvements in overall performance. The addition of either RGB or segmentation conditioning lead to a similar scores of $0.37$ and $0.36$ respectively. Overall, the synthetic samples from SurgiFlowVid prove very beneficial for both the balanced and the under-represented classes.

\begin{table}[!htbp]
\caption{\textbf{Surgical action recognition} on the SAR-RARP$50$ dataset. Segmental F1 scores are reported. } 
\label{tab:sar_tab_segf1}
\begin{center}
{\small
\resizebox{\linewidth}{!}{
\begin{tabular}{l cc cccccc|c}
\toprule
\multirow{2}{*}{Training data} & \multicolumn{2}{c}{Cond. type} &
\cellcolor{lightmaroon!20}\makecell{Pick\\the needle} & 
\makecell{Position\\the needle} & 
\makecell{Push\\the needle} & 
\makecell{Pull\\the needle} & 
\cellcolor{lightmaroon!20}\makecell{Cut\\the suture} & 
\cellcolor{lightmaroon!20}\makecell{Return\\the needle} & Mean. \\
\cmidrule(lr){2-3}
 & Text & Sparse mask & & & & & & &\\
\midrule
Real & -- & -- & $0.28_{\pm0.17}$ & ${0.40_{\pm0.16}}$ & \ssbest{$0.62_{\pm0.18}$} & $0.41_{\pm0.14}$ & $0.09_{\pm0.08}$ & $0.22_{\pm0.18}$ & $0.32_{\pm0.06}$\\ \midrule
Real $+$ Endora & -- & -- & $0.23_{\pm0.13}$ & $0.38_{\pm0.06}$ & $0.55_{\pm0.09}$ & $0.41_{\pm0.10}$ & $0.09_{\pm0.09}$ & $0.21_{\pm0.08}$ & $0.31_{\pm0.08}$\\
Real $+$ SurV-Gen (w/o RS) & \cmark & -- & $0.26_{\pm0.12}$ & $0.40_{\pm0.04}$ & $0.55_{\pm0.08}$ & $0.41_{\pm0.06}$ & $0.12_{\pm0.09}$ & $0.23_{\pm0.12}$ & $0.33_{\pm0.04}$\\
Real $+$ SurV-Gen (RS) & \cmark & -- & $0.27_{\pm0.14}$ & $0.40_{\pm0.15}$ & $0.58_{\pm0.19}$ & \ssbest{$0.42_{\pm0.18}$} & \best{$0.20_{\pm0.13}$} & $0.23_{\pm0.18}$ & $0.35_{\pm0.07}$\\
Real $+$ SparseCtrl & \cmark & RGB & \best{$0.32_{\pm0.20}$} & \ssbest{$0.41_{\pm0.16}$} & $0.57_{\pm0.17}$ & $0.44_{\pm0.15}$ & $0.10_{\pm0.09}$ & \ssbest{$0.25_{\pm0.11}$} & \ssbest{$0.35_{\pm0.03}$} \\
\midrule
Real $+$ SurgFlowVid & \cmark & -- & $0.27_{\pm0.14}$ & $0.40_{\pm0.16}$ & $0.57_{\pm0.16}$ & $0.43_{\pm0.13}$ & $0.13_{\pm0.08}$ & $0.16_{\pm0.07}$ & $0.33_{\pm0.04}$\\
Real $+$ SurgFlowvid & \cmark & RGB & \ssbest{$0.31_{\pm0.17}$} & \best{$0.43_{\pm0.17}$} & \ssbest{$0.59_{\pm0.16}$} & \best{$0.45_{\pm0.10}$} & \ssbest{$0.15_{\pm0.04}$} & \best{$0.31_{\pm0.12}$} & \best{$0.37_{\pm0.03}$} \\
\bottomrule
\end{tabular}
}}
\end{center}
\end{table}

\begin{table}[!htbp]
\caption{\textbf{Surgical action recognition} on the SAR-RARP$50$ dataset. Segmental F1 scores are reported. for seg. frame conditioning. } 
\label{tab:sar_tab_segf1_seg}
\begin{center}
{\small
\resizebox{\linewidth}{!}{
\begin{tabular}{l cc cccccc|c}
\toprule
\multirow{2}{*}{Training data} & \multicolumn{2}{c}{Cond. type} &
\cellcolor{lightmaroon!20}\makecell{Pick\\the needle} & 
\makecell{Position\\the needle} & 
\makecell{Push\\the needle} & 
\makecell{Pull\\the needle} & 
\cellcolor{lightmaroon!20}\makecell{Cut\\the suture} & 
\cellcolor{lightmaroon!20}\makecell{Return\\the needle} & Mean. \\
\cmidrule(lr){2-3}
 & Text & Sparse mask & & & & & & &\\
\midrule
Real & -- & -- & $0.28_{\pm0.17}$ & ${0.40_{\pm0.16}}$ & \ssbest{$0.62_{\pm0.18}$} & $0.41_{\pm0.14}$ & $0.09_{\pm0.08}$ & $0.22_{\pm0.18}$ & $0.32_{\pm0.06}$\\ 
Real $+$ SparseCtrl & \cmark & Seg & \best{$0.33_{\pm0.19}$} & \best{$0.43_{\pm0.14}$} & \ssbest{$0.60_{\pm0.19}$} & \best{$0.44_{\pm0.15}$} & \ssbest{$0.12_{\pm0.10}$} & \ssbest{$0.20_{\pm0.10}$} & \ssbest{$0.35_{\pm0.05}$} \\
Real $+$ SurgFlowvid & \cmark & Seg & \ssbest{$0.30_{\pm0.14}$} & \ssbest{$0.42_{\pm0.16}$} & $0.58_{\pm0.14}$ & \ssbest{$0.43_{\pm0.13}$} & \best{$0.13_{\pm0.08}$} & \best{$0.32_{\pm0.11}$} & \best{$0.36_{\pm0.02}$} \\
\bottomrule
\end{tabular}
}}
\end{center}
\end{table}

\subsection{Ablation on sparse frames} \label{sec:sparse_frame}
\begin{figure}[!htbp]
    \centering
    \includegraphics[width=\textwidth, keepaspectratio]{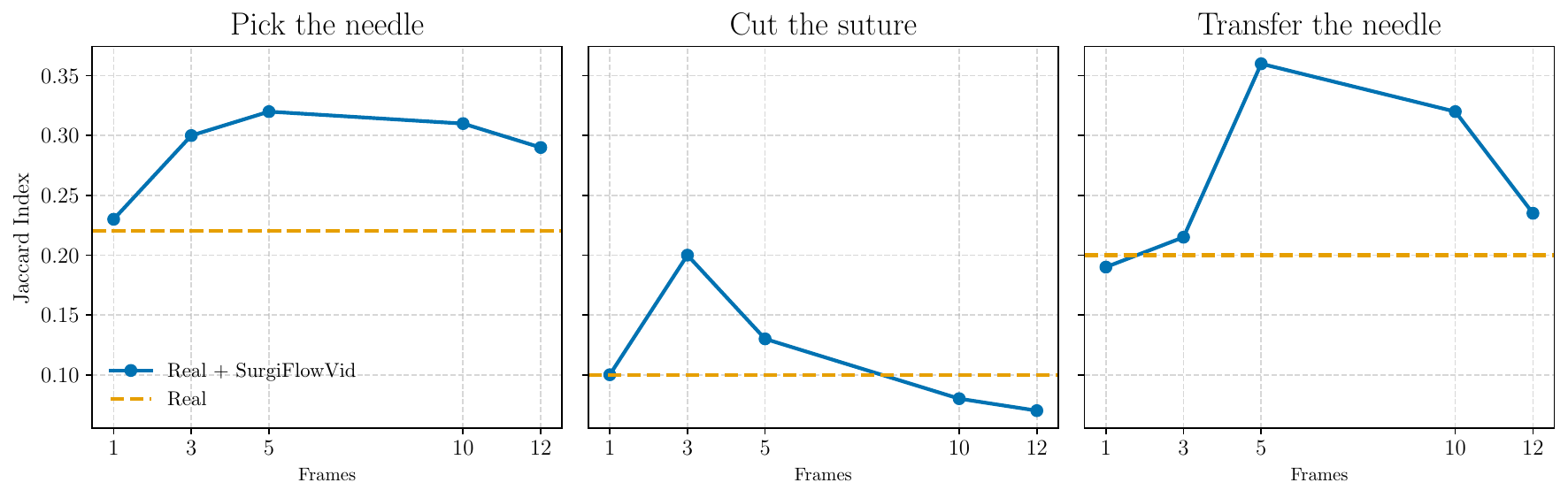}
    \caption{\textbf{Frame ablation}. The ablation on the number of sparse RGB frames on the SAR-RARP$50$ dataset. The results consists of using a X$3$D model with all the minor classes modeled together. }
    \label{fig:abl1}
\end{figure}

\begin{figure}[!htbp]
    \centering
    \includegraphics[width=\textwidth, keepaspectratio]{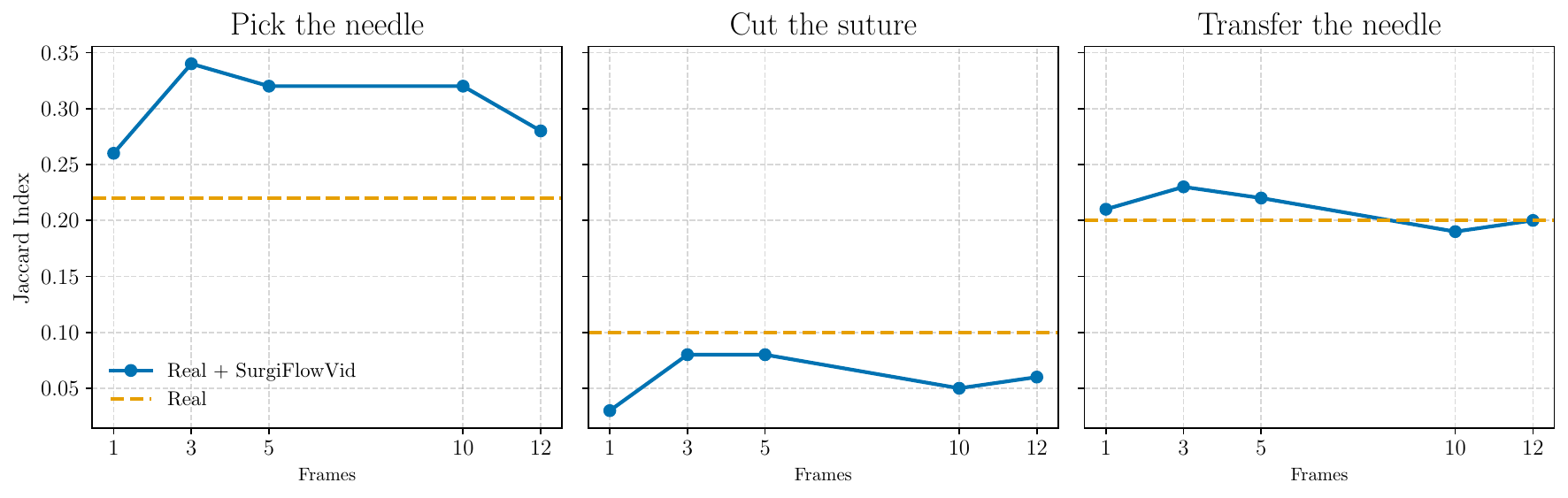}
    \caption{\textbf{Frame ablation}. The ablation on the number of sparse RGB frames on the SAR-RARP$50$ dataset. The results consists of using a X$3$D model with all the minor classes modeled separately.}
    \label{fig:abl2}
\end{figure}

We conducted an ablation study to examine the effect of the number of sparse RGB frames used during generation. We hypothesized that too few frames would provide insufficient controllability, while too many would replicate training data, reducing diversity. To test this, we varied the number of conditioning frames ($1,3,5,10,12$) and generated videos, comparing their performance against models trained solely on real data. Results are shown in Fig.~\ref{fig:abl1} (all minor classes modeled jointly) and Fig.~\ref{fig:abl2} (each class modeled separately). A consistent trend across both settings is that using only one frame yields performance similar to the real-only baseline, indicating limited consistency and, in some cases, degenerate generations. Conversely, conditioning on 12 of the 16 frames produced results close to the real dataset baseline, as little additional diversity was introduced. Based on these findings, we adopted a strategy of sampling 3–5 random frames from the real dataset as conditional inputs. These experiments were initially conducted with the X3D model, and the same frame distribution was subsequently applied across all experiments, including the SparseCtrl baseline.

\subsection{Model Analysis} \label{sec:model_analysis}
In this section, we analyze the model in terms of the video generation cost. The results are shown in Tab.~\ref{tab:model_analysis}. In comparison to Endora, both SurgV-Gen and our approach have lesser number of training parameters as the training is conducted in different stages. Our method, SurgiFlowVid is capable of generating videos at the resolution of $512\times512$ pixels whereas the baselines, SurV-Gen and SparseCtrl generates videos at $256\times256$ pixels and Endora at $128\times128$ pixels. We also train our approach at $512\times512$ pixels. Our framework is capable of training at lower resolutions but we opted to train them at higher resolutions as it could be helpful for the downstream task. There exists certain organs or auxillary tool structures which appears to be very small in shape. Generating videos at higher resolution can benefit these downstream models to learn these spatial structures effectively. We noticed the benefits for the classification of \emph{catheter} and \emph{clamps} in SAR-RARP$50$ dataset with synthetic videos from SurgiFlowVid (see Tab.~\ref{tab:tool_sar}). However, an analysis on the video resolution for the downstream task could shed more insights and we leave that for future work. As we generate videos at higher resolution, our approach requires a small overhead in terms of training and sampling times. We believe with the innovations in high performant GPUs these costs could be lowered drastically.

\begin{table}[!htbp]
\caption{\textbf{Model analysis}. The various parameters of the different baselines. SVE denotes the sparse visual encoder in our approach. The inference time was measured on a A100-40GB GPU.} 
\label{tab:model_analysis}
\begin{center}
{\small
\resizebox{0.85\linewidth}{!}{
\begin{tabular}{lcccc}
\toprule
Method & \makecell{Trainable\\params. (M)} & \makecell{Video\\resolution} & \makecell{Sampling\\steps} & \makecell{Inf.\\time(sec)} \\ 
\midrule
Endora & $675$ & $128\times128$ & $50$ & $7.85$s \\
SurV-Gen & $435$ & $256\times256$ & $50$ & $6.55$s \\
SurgiFlowVid & $437$ & $512\times512$ & $50$ & $7.53$s \\
SparseCtrl & $453$ & $256\times256$ & $30$ & $10.20$s \\
SurgiFlowVid + SVE & $456$ & $512\times512$ & $30$ & $10.45$s  \\
\bottomrule
\end{tabular}
}}
\end{center}
\end{table}

\begin{table}[!htbp]
\caption{\textbf{Image quality metrics}. The CLIP image score of different methods are reported here. Higher is better.} 
\label{tab:model_image_clip}
\begin{center}
{\small
\resizebox{0.85\linewidth}{!}{
\begin{tabular}{lccccccccc}
\toprule
Method & \multicolumn{3}{c}{SAR-RARP$50$} & \multicolumn{2}{c}{GynSurg} & \multicolumn{4}{c}{GraSP} \\
\midrule
 & A$1$ & A$5$ & A$7$ & P$3$ & P$4$ & G$1$ & G$2$ & G$3$ & G$4$ \\
 \midrule
Endora & $70.30$ & $66.85$ & $73.65$ & $69.43$ & $70.12$ & $57.09$ & $68.10$ & $74.41$ & $60.72$ \\
SurV-Gen & $75.30$ & $70.22$ & $78.85$ & $71.30$ & $75.83$ & $62.15$ & $73.10$ & $68.32$ & $62.15$ \\
SurgiFlowVid & $74.46$ & $76.08$ & $78.25$ & $72.95$ & $66.76$ & $68.20$ & $70.10$ & $72.15$ & $65.27$ \\
\bottomrule
\end{tabular}
}}
\end{center}
\end{table}

\subsection{Image Quality Analysis} \label{sec:image_quality}
 As our goal is to mitigate data imbalance, we focused primarily on generating videos of under-represented classes and evaluating them on the downstream task. We consider this approach as an effective way to directly measure the effectiveness and the usefulness of the synthetic videos. In this section, we evaluate the quality of the generated videos with  the CLIP~\citep{hessel2021clipscore} image and the LPIPS~\citep{zhang2018unreasonable} score. Both these metrics evaluate the quality of the generated frames using features from pre-trained models on large-scale natural images. The results are shown in Tab.~\ref{tab:model_image_clip} and Tab.~\ref{tab:model_image_lpips}. We compare our approach, SurgiFlowVid with text conditioning against Endora and SurV-Gen. We do not compute these scores for SparseCtrl or sparse visual encoder using our approach, as there already exists frames from the real dataset. The image quality varied between different classes and we did not notice a co-relation between these scores to the downstream model performance. Hence, these values should be interpreted with caution given that they are computed with pre-trained weights from models not trained on surgical images/videos.

\begin{table}[!htbp]
\caption{\textbf{Image quality metrics}. The LPIPS score of different methods are reported here. Lower is better.} 
\label{tab:model_image_lpips}
\begin{center}
{\small
\resizebox{0.85\linewidth}{!}{
\begin{tabular}{lccccccccc}
\toprule
Method & \multicolumn{3}{c}{SAR-RARP$50$} & \multicolumn{2}{c}{GynSurg} & \multicolumn{4}{c}{GraSP} \\
\midrule
 & A$1$ & A$5$ & A$7$ & P$3$ & P$4$ & G$1$ & G$2$ & G$3$ & G$4$ \\
 \midrule
Endora & $0.70$ & $0.53$ & $0.59$ & $0.54$ & $0.49$ & $0.63$ & $0.66$ & $0.65$ & $0.63$ \\
SurV-Gen & $0.68$ & $0.54$ & $0.57$ & $0.51$ & $0.56$ & $0.57$ & $0.67$ & $0.71$ & $0.74$ \\
SurgiFlowVid & $0.66$ & $0.56$ & $0.52$ & $0.49$ & $0.50$ & $0.51$ & $0.60$ & $0.74$ & $0.72$ \\
\bottomrule
\end{tabular}
}}
\end{center}
\end{table}


\subsection{Laparoscope motion} \label{sec:lap_motion}
\begin{figure}[!htbp]
  \centering
  \includegraphics[width=\linewidth]{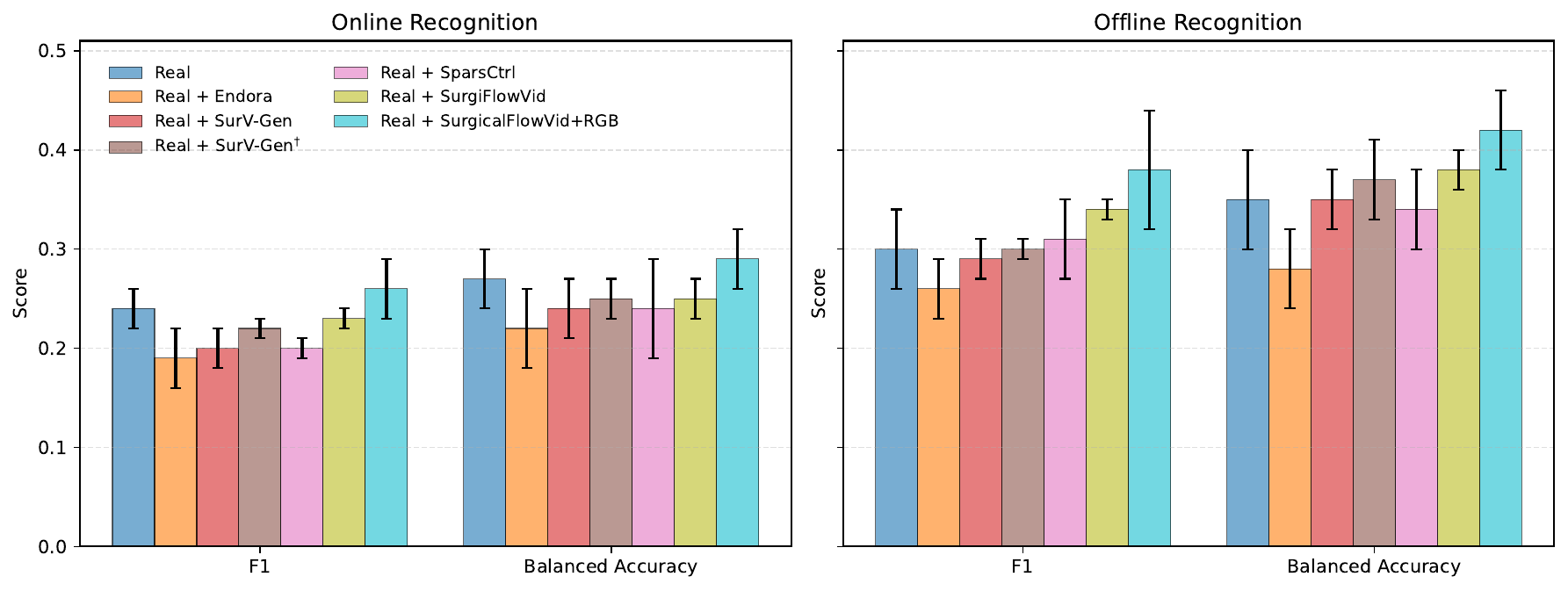}
  \caption{Laparoscope motion prediction on the Autolaparo dataset.
  Bars show mean score with standard deviation (error bars). }
  \label{fig:auto_comp_suppl}
\end{figure}

In addition to the F$1$ score, we also computed the balanced accuracy as an additional metric. Fig.~\ref{fig:auto_comp_suppl} shows the results on the laparoscope motion prediction task. Similar to the results seen in Fig.~\ref{fig:auto_comp}, the overall scores are higher for the the offline recognition. 

\section{Dataset} \label{sec:dataset_add}
\underline{SAR-RARP$50$}: The dataset consists actions annotated at $10$ fps. Our initial experiments indicated this temporal frame to be very fine and hence we chose to sample the frames at $5$ fps. The annotations for the surgical tools were available at $1$ fps making it sparse in nature. For the sparse conditional generation, we randomly samples video frames in the range $3-5$ and place them in a different temporal order than the real dataset, so as to create the synthetic data as diverse as possible. For the sparse segmentation conditioning, we opted to include a minimum of $4$ frames in the $16$ frames video clips during training and sampling time. \\
\underline{GraSP}: This dataset consists of annotations at both $30$ and $1$ fps temporal windows. As $1$fps was very coarse in nature, we opted to sample frames at $5$ fps from the $30$ fps annotations. The segmentation annotations were available at every $35$ seconds making them very sparse in nature. Based on dataset analysis, we noticed that creating video clips with at least one segmentation frame as conditioning for the under-represented samples were very challenging. Hence, we opted out of segmentation frames conditioning for the sparse visual encoder in our experiments. However, for the surgical tool presence detection task, we sampled a minimum of $4$ frames around the available segmentation frame and used it as the conditioning to generate videos for this task. 

The details on the addition of synthetic samples are shown in Tab.~\ref{tab:data_analysis}.

\begin{table}[!htbp]
\caption{\textbf{Dataset details}. The values in the table include the total number of video clips from the training set. We add only synthetic samples to the under-represented classes to match and balance the instances with the well balanced classes. } 
\label{tab:data_analysis}
\begin{center}
{\small
\resizebox{0.65\linewidth}{!}{
\begin{tabular}{lccc}
\toprule
Dataset & Step/action class & \makecell{Data points\\in real dataset} & \makecell{Added\\syn. samples} \\
\midrule
\multirow{6}{*}{SAR-RARP$50$} & Pick the needle & $332$ & $900$ \\ 
& Position the needle & $1329$ & - \\ 
& Push the needle & $1395$ & -  \\ 
& Pull the needle & $1208$ & - \\ 
& Cut the suture & $115$ & $1100$\\ 
& Return the needle & $168$ & $1100$ \\ 
\midrule
\multirow{5}{*}{GraSP} & Pull the suture & $992$ & $1600$ \\ 
& Tie the suture & $712$ & $1800$ \\ 
& Cut the suture & $1213$ & $1300$ \\ 
& \makecell{Cut btw. \\the prostate} & $1616$ & $1000$ \\ 
& Identify iliac artery & $2800$ & - \\ 
\midrule
\multirow{4}{*}{GynSurg} & Coagulation & $690$ & -\\ 
& Needle passing & $869$ & - \\ 
& Suction/Irrigation & $267$ & $550$\\ 
& Transection & $168$ & $650$\\ 
\bottomrule
\end{tabular}
}}
\end{center}
\end{table}

\section{model training} \label{sec:diff_train}
\subsection{Diffusion Image pre-training} \label{sec:diffusion_image}
We build upon the SurV-Gen model~\citep{venkatesh2025mission}, which was initially proposed to generate synthetic samples of under-represented classes to mitigate data imbalance in surgical datasets. The framework adopts a multi-stage training procedure. In the first stage, frames are extracted from the training split of surgical videos and a $2$D Stable Diffusion (SD) model~\citep{sd} is trained. We follow the same pipeline with several modifications. Training the spatial SD directly on the limited frames from the downstream task datasets can result in overfitting, reduced diversity of generated frames, or potential data leakage. This phenomenon was observed in SurV-Gen, where synthetic augmentation yielded only marginal improvements without rejection sampling.  

To address this issue, we curated an in-house dataset comprising video recordings from different surgical procedures. The dataset consists of approximately $7000$ clips, each ranging from $6$ to $8$ minutes in length. From this collection, we extracted $\sim 4000$ frames to train the $2$D component of the model. We initialized training from the SD-v$1.5$ checkpoint, pre-trained on the large-scale LAION-5B dataset~\citep{laion}, which provided a strong initialization compared to training from scratch. The model was fine-tuned for $3000$ steps using the AdamW optimizer~\citep{loshchilov2017decoupled} with a learning rate of $1\text{e}^{-4}$, a batch size of $2$, and gradient checkpointing enabled. Due to computational constraints, frames were resized from their original resolution of $1048 \times 2048$ to $512 \times 512$. For text conditioning, we employed simple prompts such as \textit{``An image of a surgical procedure''}, with embeddings generated using the CLIP text encoder~\citep{learning}. This fine-tuned SD model served as the base $2$D diffusion prior for any subsquent $2$D diffusion models. We fine-tune this model on the downstream datasets before video diffusion training. The spatial priors are learnt during this stage.

\subsection{Diffusion Video pre-training} \label{sec:diffusion_pretrainin}

Next, we focus on the video training stage. In the SurV-Gen approach, the spatial layers are frozen and only the temporal attention layers are trained during the second stage. In contrast, our framework trains the temporal layers jointly with both RGB and optical flow frames. To further improve temporal modeling, we investigated a video pre-training strategy inspired by previous works on video diffusion models~\citep{svd, moviegen}. Our hypothesis is that temporal motion priors, such as the movement of tools, tissue motions andpartially tool tissue interactions can be better learned by training on the unconditional internally curated dataset, which contains diverse anatomical structures, varying illumination conditions, different endoscope motions, and a wide range of surgical tools and tool interactions. This dataset introduces substantial variability that more closely reflects real-world surgical scenarios.  

To test this, we extended SurV-Gen and trained it in two ways, keeping the training recipe unchanged (i.e., only the temporal attention layers are updated). First, we trained SurV-Gen directly on the SAR-RARP$50$ dataset, where the $2$D SD backbone was also trained on frames extracted from the same dataset. Second, we replaced the $2$D SD backbone with our fine-tuned $2$D model and pre-trained the temporal layers on the curated dataset of $\sim7000$ videos. For this, we created overlapping subsets of $3000$, $5000$, and $7000$ videos, each containing at least $1500$ new clips. The pre-trained temporal layers were then fine-tuned on SAR-RARP$50$. 

\begin{wrapfigure}{r}{0.5\textwidth}  
  \centering
  \includegraphics[width=0.45\textwidth]{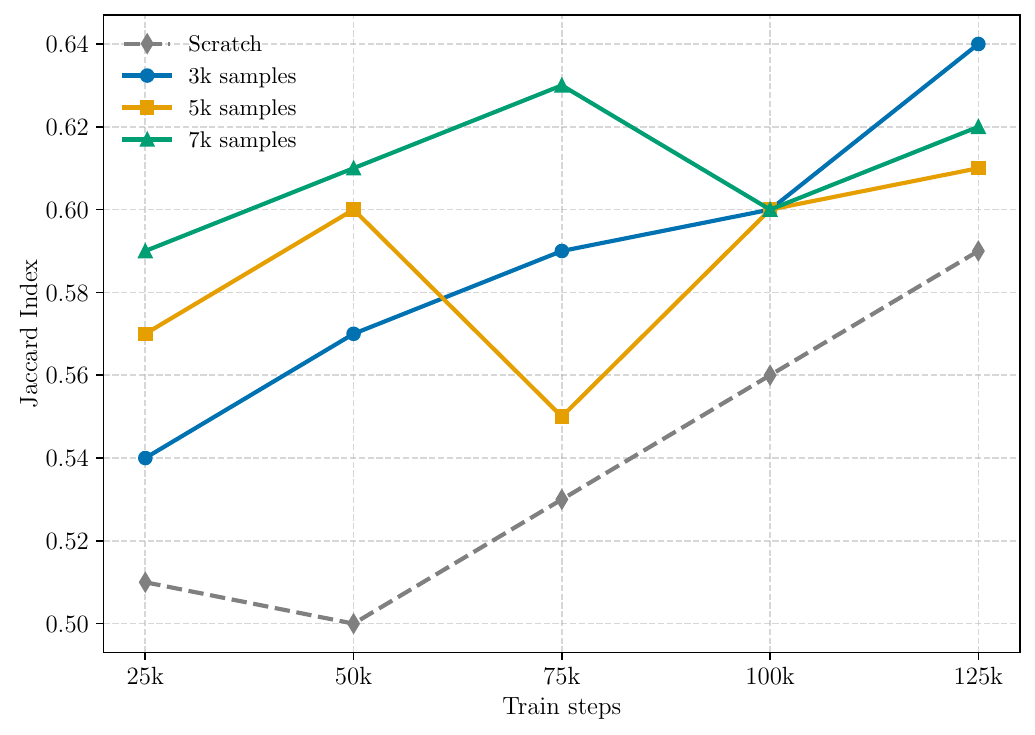} 
  \caption{The results on video pre-training.}
  \label{fig:pretrain}
\end{wrapfigure}

This pre-training strategy is expected to accelerate learning of spatio-temporal representations from the limited SAR-RARP$50$ data. We then generated synthetic samples of under-represented classes using label guidance, following SurV-Gen, and evaluated their impact on downstream action recognition performance. The results are shown in Fig.~\ref{fig:pretrain}. 

We analyzed only the three under-represented classes and report the weighted average Jaccard index of these classes. We notice that the pre-training strategy leads to higher recognition scores in comparison to using only the real dataset for the same number of training steps. We noticed smaller dips in performance for the $5$k and $7$k samples, which could be attributed to a distributional shift to the SAR-RARP$50$ dataset. On the other hand, we noticed a continuous improvement in jaccard scores for the $3$k samples. Overall, these results indicate that the pre-training strategy leads to learning the spatio-temporal relationships better, such that when minimal data is available, the model can learn faster. Based on these results, we used the $2$D spatial SD model and temporal attention layers pre-trained on our internal dataset as the starting checkpoints for the SurgiFlowVid training scheme.  

\subsection{SurgiFlowVid training} \label{sec:surgiflow_train}
 Based on these results we opted to use the temporal layers trained on our internal dataset as the pre-trained model. This offers the advantage that, the SurgiFlowVid training time reduces and also we can avoid the over-fitting of the dataset given the fact that there exists only limited training data from the downstream datasets. We fine-tune the pre-trained temporal attention layers using our proposed dual-prediction U-net module. The optical flow frames are extracted using the RAFT model~\citep{raft}. For SurgiFlowVid training, we extract clips of $16$ frames at a frame rate of $5$ for all the datasets. The hyperparameter details are mentioned in Tab.~\ref{tab:hyper}.

 \subsection{Downstream model training} \label{sec:downstream}
For the action recognition task (SAR-RARP$50$), we used the MviT-v2 model from the SlowFast library\footnote{\url{https://github.com/facebookresearch/SlowFast}}. We downsampled the videos to $224\times384$ pixels for training with a temporal resolution of $5$ fps. Image augmentations such as PCA jitter, RGB scale shift, brightness and contrast shift, random flipping with scale cropping was used along with inverse frequency balancing during the training on the real data. For additional details on the model, readers can refer to SlowFast repo. We followed the similar recipe for the GynSurg dataset. The model were trained for $150$ epochs with a learning rate of $1e^{4}$ with the best model being chosen using a validation dataset.

For the GraSP dataset, we used the similar settings from the TAPIS model\footnote{\url{https://github.com/BCV-Uniandes/GraSP/tree/main/TAPIS}}. It is to be noted that we do not compare the values directly to the work from~\citep{grasp} on the GraSP dataset. This is due to the fact that the results reported from the TAPIS model have been obtained directly using the test set as the selection criteria during training. We create a separate validation set from the training set which we use as the selection criteria of the trained model. The test set is clearly separated during the training of both diffusion and downstream models to avoid any data leakage. For the combined training of real and synthetic videos, we opted for a simple and easier strategy than rejection sampling as proposed in~\citep{venkatesh2025mission}. We sampled a batch of data points such that $25\%$ of this batch consists of synthetic videos. We chose this method as it works on the fly during training and the time and effort in rejecting synthetic samples are drastically reduced. 

For the surgical tool presence detection task, we used the Swin transformer model. The videos were resized to a resolution of $384\times384$ during training with augmentations such as RGB channel shift, scaled cropping and temporal shift. We trained the models using binary cross entropy loss with weighted sampling to include the imbalance in the surgical tools.

\begin{table}[!htbp]
    \centering
    \begin{tabular}{lccc}
         Hyperparameter & Image fine-tuning & Video-pretraining & SurgiFlowVid training \\
         \midrule
         \midrule
         \textbf{Datatset} & & & \\
         No. of samples & $4000$ & $7000$ & Train split of the dataset\\
         Resolution & $512\times512$ & $256\times256$ \& $512\times512$  & $512\times512$\\
         Video length & - & $16$ frames & $16$ frames \\
         Sample rate & - & $5$ & $4$-$5$\\
         Context length & - & $16$ & $16$ \\
         \midrule
         \midrule
         \textbf{Model params} & & & \\
         Pre-trained model & SDv-$1.5$ & \makecell{Pre-trained on \\internal} & \makecell{Pre-trained on \\internal} \\
         Params frozen & - & Spatial layers & Spatial layers \\
         \midrule
         \midrule
         \textbf{Temporal layers} & & & \\
         Depth &  - & $2$ & $2$ \\
         Temporal resolution & - & $[1,2,4,8]$ & $[1,2,4,8]$ \\
         Head channels &  - & $16$ & $16$ \\
         No. of heads &  - & $8$ & $8$ \\
         Position encoding &  - & sinusoidal & sinusoidal \\
         PE dim &  - & $24$ & $24$ \\
         Cross attention dim &  - & $32$ & $32$ \\
         Act.function&  - & GeLU & GeLU \\
         \midrule
         \midrule
         \textbf{Training params} & & & \\
         Optimizer & AdamW & AdamW & AdamW \\
         Learning rate & $1e^{-4}$ & $1e^{-5}$ & $1e^{-5}$ \\
         Lr warm steps & $500$ & $5000$ & $5000$ \\
         Lr scheduler & cosine & cosine & cosine \\
         $\beta_{1}$ & $0.9$ & $0.9$ & $0.94$ \\
         $\beta_2$ & $0.999$ & $0.999$ & $0.995$ \\
         Weight decay $\omega$ & $1e^{-2}$ & - & - \\
         Train steps & $3000$ & $125$k & $75-125$k \\
         \midrule
         \midrule
         \textbf{Train timestep} & & & \\
         Diffusion step & $1000$ & $1000$ & $1000$ \\
         Noise schedule & linear & linear & linear \\
         $\beta_{0}$ & $1e-^{4}$ & $0.00085$ & $0.00085$ \\
         $\beta_{T}$ & $0.02$ & $0.012$ & $0.012$ \\
         \midrule
         \midrule
         \textbf{Sampling params} & & & \\
         Sampler & DDPM & DDIM & DDIM \\
         Steps & - & 50 & 50 (30 for SVE) \\
         CFG scale & $6.5$ & $5.5$ & $5.0$ \\
         \midrule
         \midrule
         \textbf{Device requirements} & & & \\
         GPU-type & A100-40GB & H200-80GB & H200-140GB \\
         No. of gpus & 1 & 1 & 1 \\ 
    \end{tabular}
    \caption{Hyperparameters for training the $2$D and the temporal attention layers of the diffusion model. SVE denotes \emph{Sparse visual encoder} used for conditional generation.}
    \label{tab:hyper}
\end{table}

\section{Qualitative Results}\label{sec:qual}
\begin{figure}[h]
\begin{center}
\includegraphics[width=\textwidth,keepaspectratio]{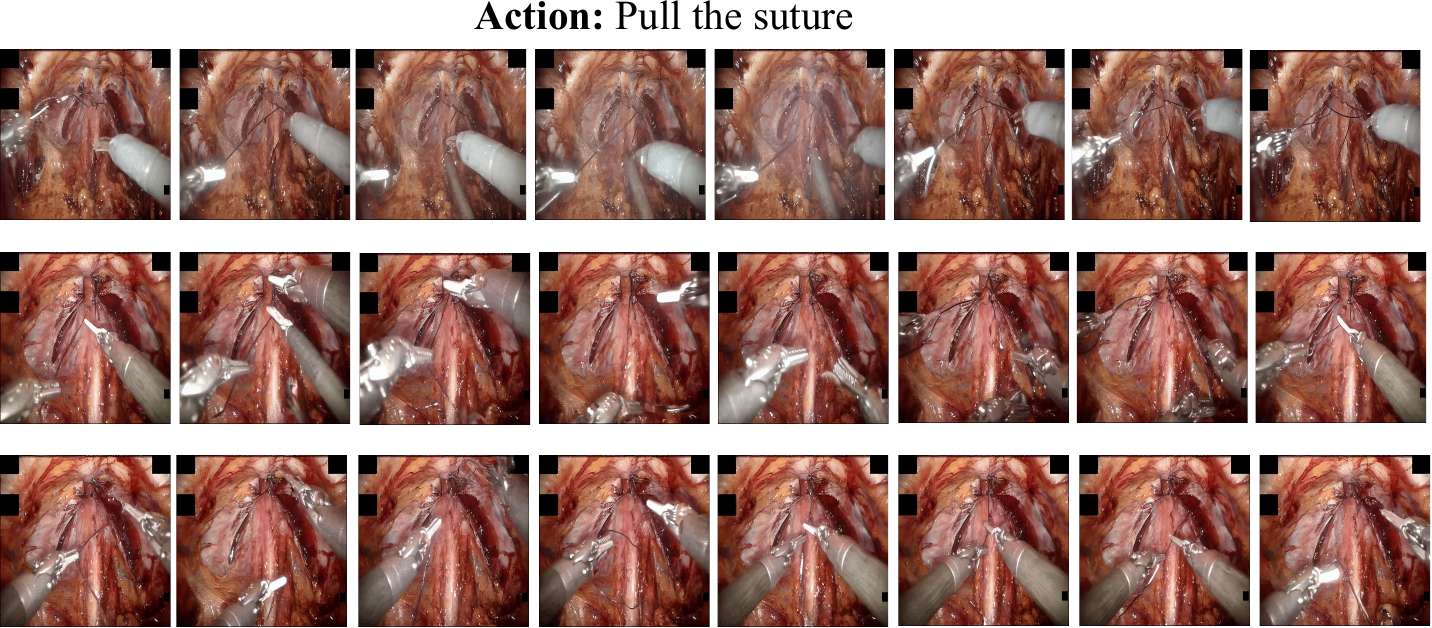}
\end{center}
\caption{Results from SurgiFlowVid with text conditioning on GraSP dataset.} 
\label{fig:gp_15}

\end{figure}
\begin{figure}[h]
\begin{center}
\includegraphics[width=\textwidth,keepaspectratio]{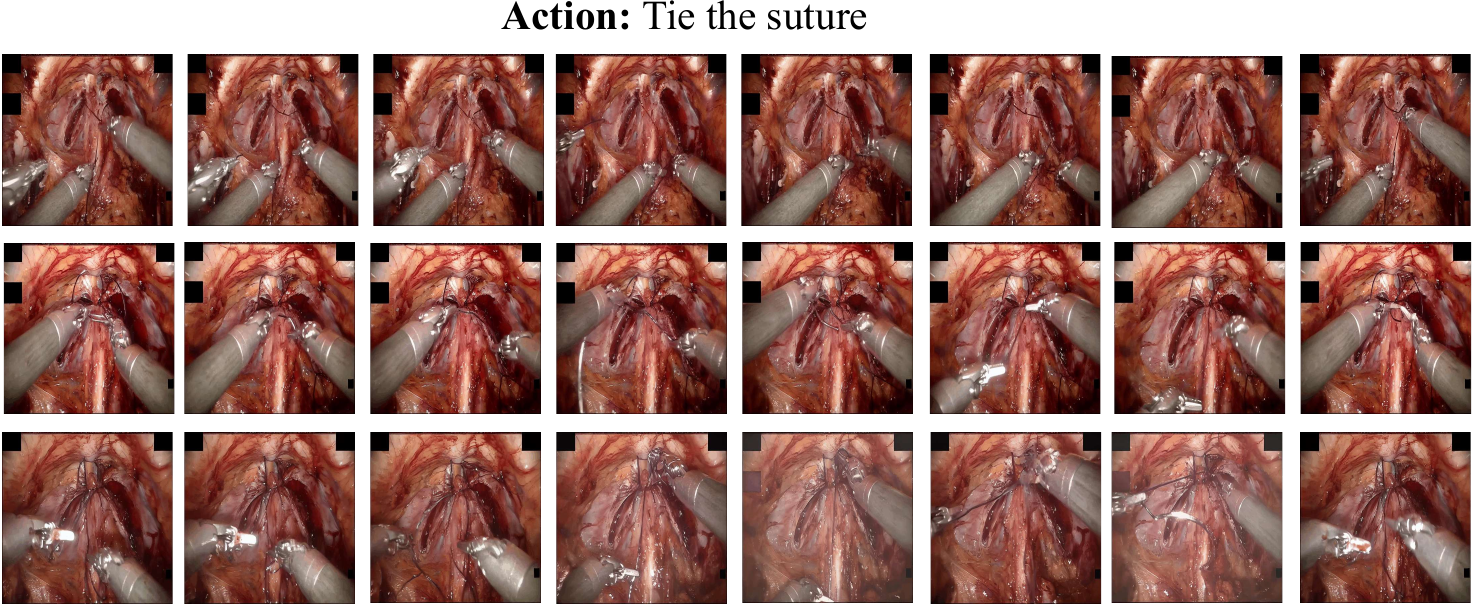}
\end{center}
\caption{Results from SurgiFlowVid with text conditioning on GraSP dataset.} 
\label{fig:gp_16}
\end{figure}

\begin{figure}[h]
\begin{center}
\includegraphics[width=\textwidth,keepaspectratio]{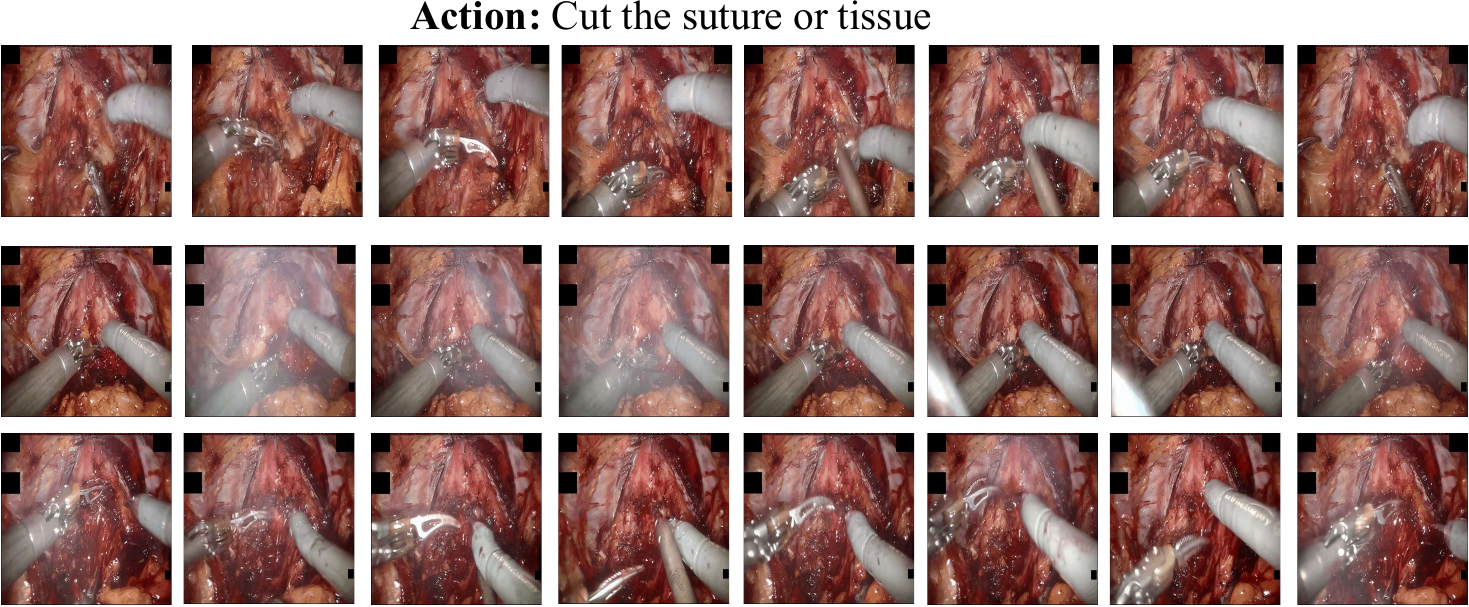}
\end{center}
\caption{Results from SurgiFlowVid with text conditioning on GraSP dataset. In the $2$nd row, we notice the presence of smoke as the tissue is cauterized using the tool.} 
\label{fig:gp_18}
\end{figure}

\begin{figure}[h]
\begin{center}
\includegraphics[width=\textwidth,keepaspectratio]{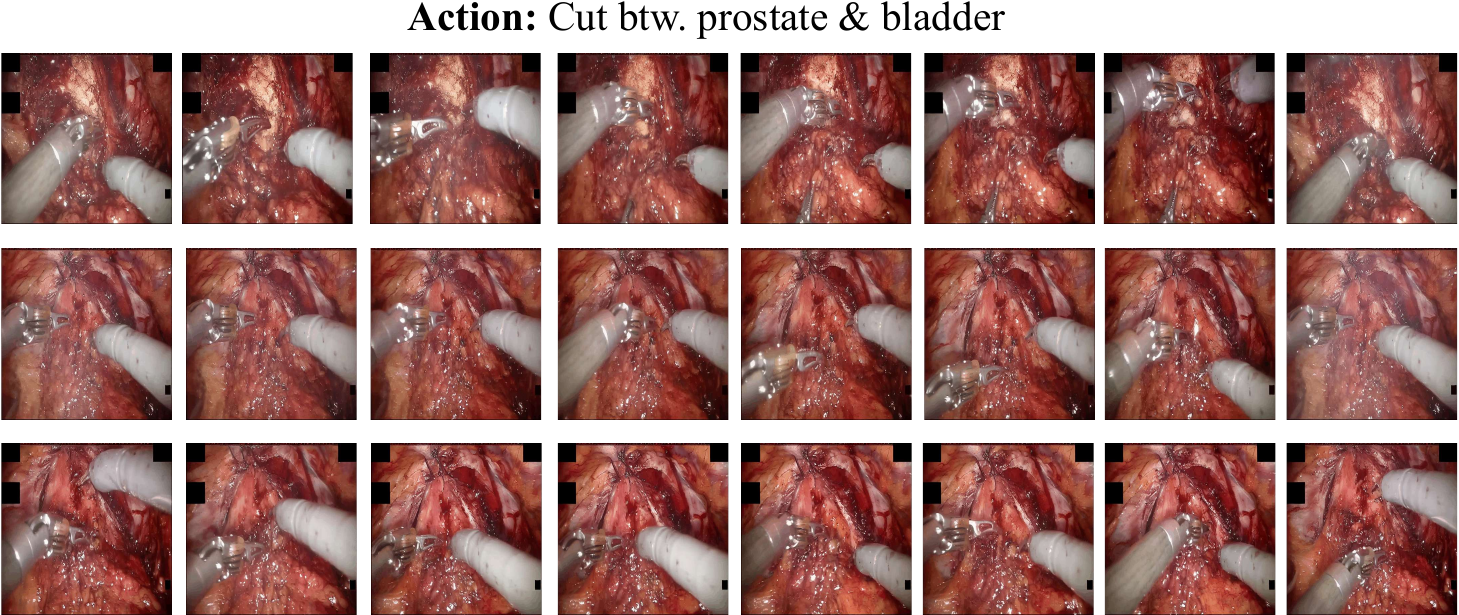}
\end{center}
\caption{Results from SurgiFlowVid with text conditioning on GraSP dataset.} 
\label{fig:gp_19}
\end{figure}

\begin{figure}[h]
\begin{center}
\includegraphics[width=\textwidth,keepaspectratio]{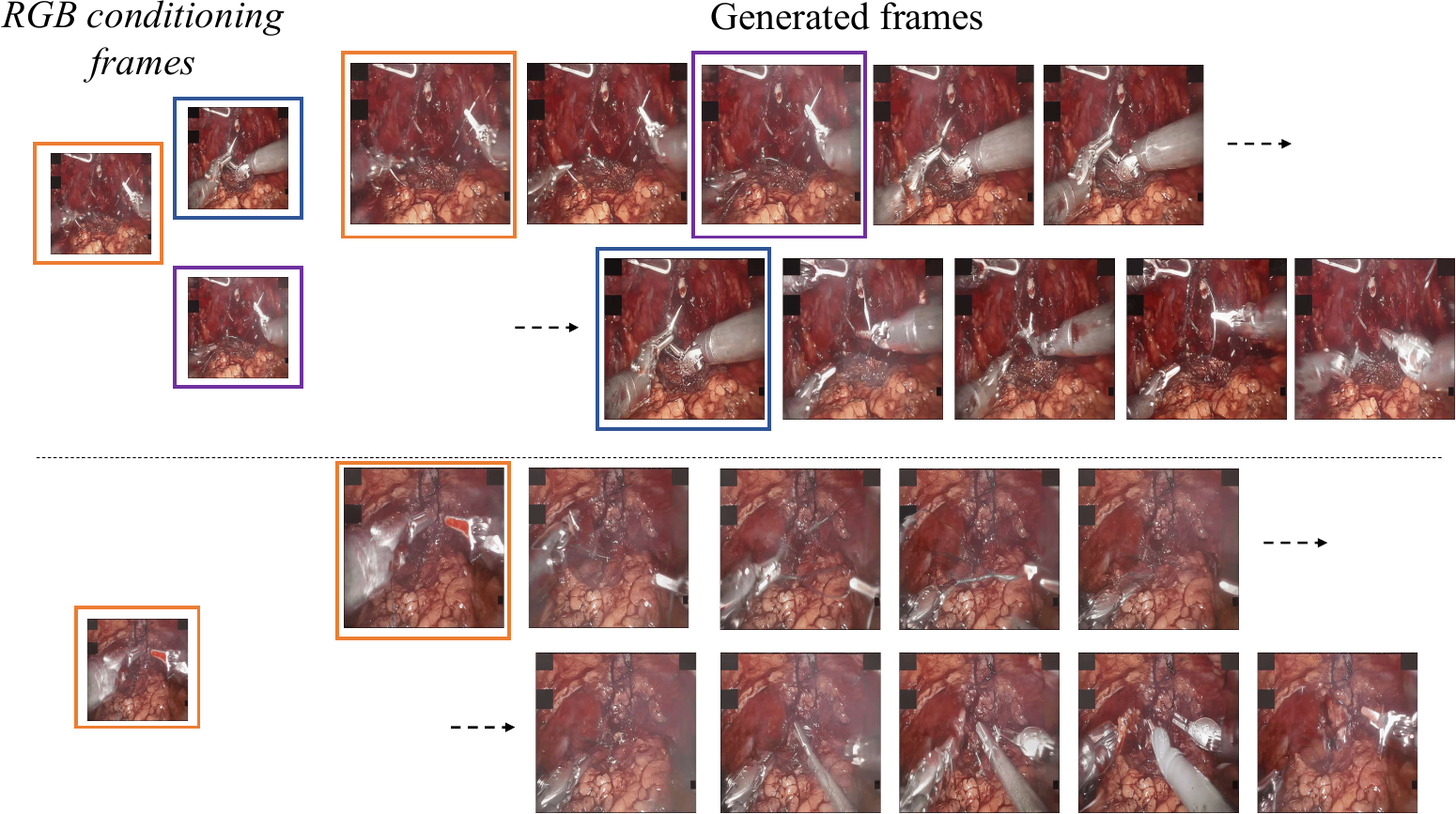}
\end{center}
\caption{Results from SurgiFlowVid with RGB conditioning on GraSP dataset.The frames on the left indicate the sparse conditioning frames and the left frames indicate the generated video frames. The coloured boxes show the position of the corresponding condition frame. The dotted arrow indicates the next subsequent frames. The action corresponds to \emph{pull the suture}.} 
\label{fig:gp_15_rgb}
\end{figure}

\begin{figure}[h]
\begin{center}
\includegraphics[width=\textwidth,keepaspectratio]{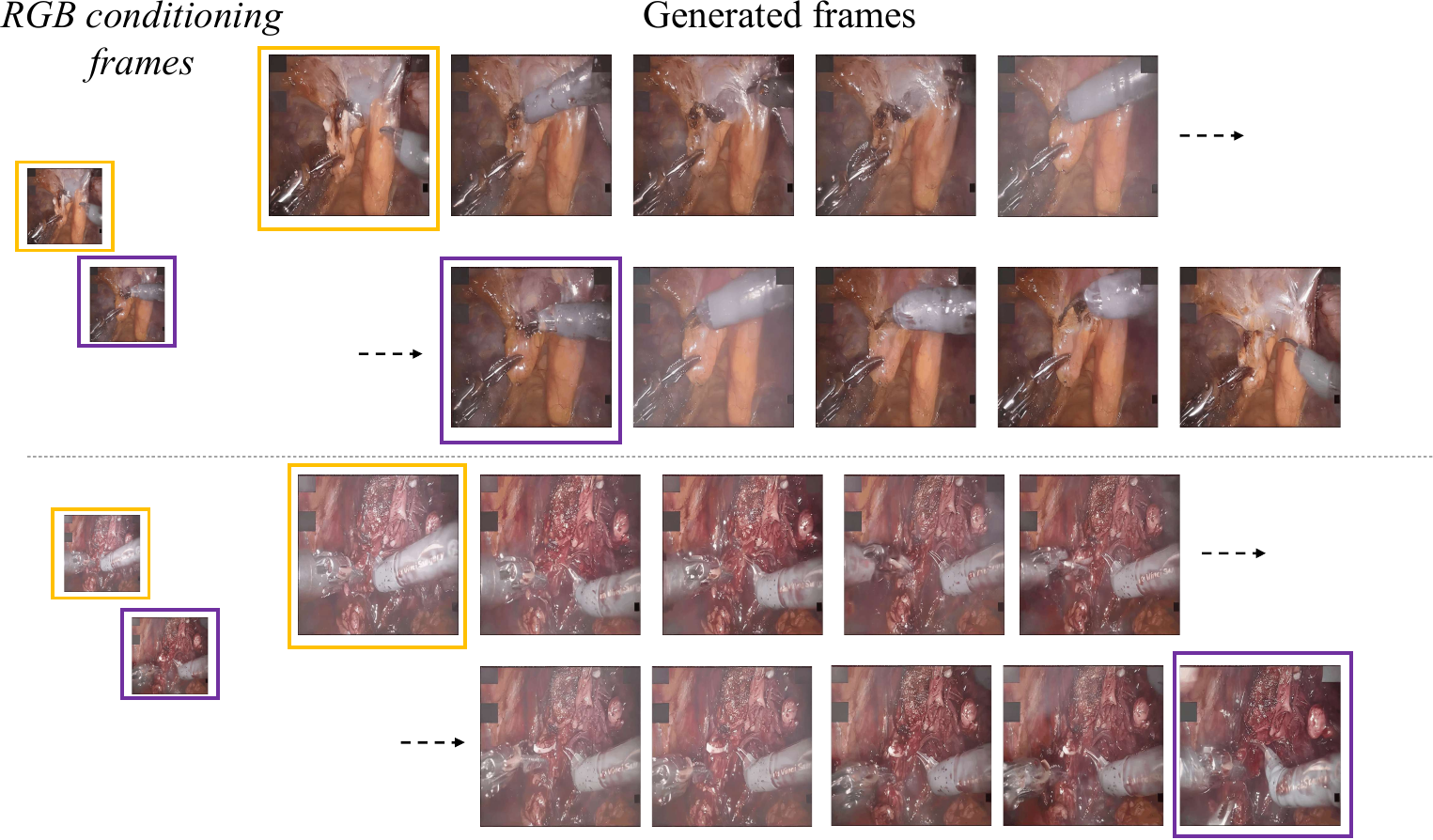}
\end{center}
\caption{Results from SurgiFlowVid with RGB frame conditioning on GraSP dataset. The action corresponds to \emph{cut the tissue}.} 
\label{fig:gp_18_rgb}
\end{figure}

\end{document}